\documentstyle[aps,epsfig]{revtex}
\newcommand{\be}{\begin{equation}}
\newcommand{\ee}{\end{equation}}
\newcommand{\bea}{\begin{eqnarray}}
\newcommand{\eea}{\end{eqnarray}}
\newcommand{\hf} {{1\over2}}
\newcommand{\nonu}{\nonumber\\}
\def\eq#1{(\ref{#1})}

\def\la{\langle}
\def\ra{\rangle}
\def\cd#1{{\cal D}[#1]}
\def\cD{{\cal D}}

\def\tcG{\tilde{\cal G}}
\def\tcGi{\tilde{\cal G}^{-1}}

\def\psid{\psi^\dagger}
\def\tr{{\mathrm Tr}}
\def\ve{V^{\mathrm ext}}
\def\ord#1{{\cal O}(#1)}
\def\dt{\Delta t}

\def\tG{\tilde G}
\def\tGi{{\tilde G}^{-1}}
\def\tS{\tilde S}
\def\rt{\rho^*}
\def\ei{{1\over e}}
\def\eik{{1\over e^2}}
\def\ide{{1\over\Delta}}
\def\mb#1{{\mathbf#1}}
\def\mr#1{{\mathrm#1}}
\def\rgr{\rho_{\mathrm{gr}}}
\def\vgr{v_{\mathrm{gr}}}
\def\cC{{\cal C}} 

\begin{document}
\title{Effective actions and the density functional theory}
\author{J. Polonyi$^{1,2}$ and K. Sailer$^3$}
\address{$^1$ Institute for Theoretical Physics, Louis Pasteur University,
Strasbourg, France}
\address{$^2$ Department of Atomic Physics, Lorand E\"otv\"os University,
Budapest, Hungary}
\address{$^3$ Department for Theoretical Physics, University of Debrecen,
Debrecen, Hungary}
\date{\today}
\maketitle
\begin{abstract}
The effective action for the charge density and the photon field
is proposed as a generalization of the density functional.
A simple definition is given for the density functional, as
the functional Legendre transform of the generator
functional of connected Green functions for the density
and the photon field, offering systematic approximation schemes.
The leading order of the perturbation expansion reproduces
the Hartree-Fock equation. A renormalization group motivated method is
introduced to turn on the Coulomb interaction gradually and to
find corrections to the Hartree-Fock and the Kohn-Sham schemes.
\end{abstract}
\section{Introduction}
The functional methods, based on the generator functionals for
the different Green functions have proven to be useful in Statistical
and Quantum Physics \cite{vasil}. There are two kinds of important generator
functionals characterizing the dynamics, corresponding to the
connected and the irreducible Green functions. They are related by
functional Legendre transformation and the latter is called the effective
action \cite{rgqft,effa}.

Another widely used application of
functionals is the density functional formalism \cite{hoko,ksh}.
In this method one introduces
an external potential and studies the response of the charge
density in the ground state \cite{geld}. The key object of this
formalism is the density functional $E_\mr{gr}[\rho]$, the ground state
energy of the system with charge density distribution $\rho(\mb{x})$.
Few simple properties based on the Hohenberg-Kohn (HK) theorems
\cite{hoko} and the use of the Kohn-Sham (KS) scheme \cite{ksh}
obtained by means of the perturbation expansion,\cite{kshpert} have
led to powerful and widely applicable approximations. The problem left open
is the determination of the density functional beyond the perturbation
expansion. 

There are two further developments in the search for the appropriate density
functional we relate in this paper. One is the so called adiabatic 
approximation to the density functional \cite{adiab} combined with the 
coordinate scaling \cite{cordsc}. In this scheme the Coulomb interaction
is turned on gradually \cite{pines} and the external potential 
is adjusted in such a 
manner that the ground state density remains invariant. The result is an
interpolation, established between a non-interactive and the
physical system. Another development is the realization
that the HK theorems emerge naturally and trivially within the framework 
of the effective action formalism \cite{hkefa}. This relation may be
found by means of the time-dependent variational method
\cite{rajag}, or by considering the Helmholtz free energy for classical 
fluids \cite{dfclfl}, as well. The perturbative emergence of the KS scheme
\cite{kseffap} is particularly simple in this language when the path
integral formalism is used \cite{kspi}.

In defining  the effective action we go beyond the scheme used in Refs.
\cite{hkefa} and \cite{kseffap} by introducing an external source coupled 
to the photon field in order to simplify some intermediate stage of the
computation. We generalize the adiabatic approximation within the
framework of the effective action formalism by borrowing ideas from the 
functional renormalization group method. The renormalization group strategy was
originally constructed to find the scale dependence in the UV region
of QED \cite{rgstp} and to treat the formal infinities in renormalizable 
Quantum Field Theories \cite{rgqft}. It was realised later that the same
method can be used to explain the singular properties of
critical systems \cite{rgcrs}. Further important applications include
the partial resummation of the perturbation expansion \cite{resum},
a systematic setting of the operator product expansion \cite{opprod}
and the solution of differential equations \cite{prdee,prdek}.
According to the original suggestion, one eliminates degrees of freedom
in the Kadanoff-Wilson blocking and traces the evolution of the reduced
dynamics as the system becomes simpler. A further
generalization of the algorithm, the renormalization group
method in the internal space allows us to turn on the
fluctuations in the path integral gradually and follow the
evolution of the dynamics in the meantime \cite{int,intcg}.
Such an implementation of the renormalization group idea is outlined
here to perform a resummation of the perturbation expansion.

The internal space renormalization group is the functional realization of the
Hellmann-Feynman theorem \cite{hel,fey} for the ground 
state and makes possible
to follow the change induced by a control parameter in
the effective action. The resulting evolution equation is simpler
than that of the adiabatic approximation. This is because the 
relation between the generator functional for the connected Green functions 
and the effective action, the Legendre transformation, makes it possible to 
realize this scheme without the readjustment of the external potential,
by turning on the Coulomb interaction alone.

Another improvement offered by the use of the functional differential
equation is the flexibility which allows us to turn the qualitative
knowledge about the dynamics into a well chosen ansatz for the effective 
action. The local density approximation \cite{hoko,ksh} supplemented
with the gradient expansion \cite{gradedf} is the usual ansatz for 
the density functional though excitations appear to bring in
non-local contributions \cite{gradepr}. 
We propose in this paper another ansatz based on the idea of the
 multi-local structure of the density functional; it is similar to the
cluster expansion in Statistical Physics.
 This ansatz
avoids the limitations of the gradient expansion and gives a systematical 
truncation scheme for the functional differential equation.
The solution of the evolution equation for the effective action
which is truncated at the bi-local level, explored in the present paper,
reproduces the HF scheme solution together with  an infinite sum 
of radiative corrections to the exchange interaction and the ground state 
energy. The efficiency of the HF method might
be explained by the fact that beyond these improvements all other
radiative corrections are generated by at least tri-local, i.e.
three-body interactions. 

A trivial lesson of the path integration method is that the
exchange interaction originates from the one-loop
level 'zero-point fluctuations' of the photon field.
The importance of this remark becomes evident
when the KS scheme is sought and one finds that the renormalization
group motivated resummation automatically generates the exchange
interactions independently of the choice of the quasi-particle wave functions.
In particular, an invariance property of the evolution equation 
with the effective action taken in the bi-local approximation
makes possible to trade the exchange terms for a local potential
energy and leads to the systematical improvement of the KS scheme.

The organization of the paper is the following.
We consider a system of electrons submerged in an
external potential and subject of the Coulomb interaction.
The generator functional $W$ for connected Green functions
of the density and the photon field and its Legendre transform
$\Gamma$, the effective action
are introduced in Section II. As the simplest example, Section
III contains a short discussion of these functionals for
non-interacting electrons. The most natural way to obtain
the effective action is perturbation expansion
in the Coulomb interaction. To perform a partial resummation
of this series we introduce in Section IV a renormalization group
motivated method, the evolution equation as the Coulomb
interaction is turned on by starting with
non-interacting electrons propagating in the physical external potential.
It is shown in Section V that the leading order contributions
of the perturbation expansion generate the HF
energy functional. The main part of this paper is
the discussion of the evolution equation for various
truncation levels for the density functional in Section VI.
Here we recover the traditional HF equations, their
systematically improved version and the generalized KS equations. Section
VI.G is devoted to a brief comparison of the effective action method
with the density functional theory. The conclusions are
given in Section VII. The Appendices present some details which
are useful in the main text, the discussion
of the physical interpretation of the effective action,
the continuum limit of the Grassmann-integral representation
of the ground state energy for fermionic systems, and Legendre
transformation as a functional differential equation.

\section{Generator functionals}
We consider in this paper fermions which have $n_s$ degenerate
one-particle spin states and the system of $N_{\mathrm{tot}}=Nn_s$ such
particles propagating in the presence of a static external potential
$\ve$ and interacting with the Coulomb potential.
After presenting the partition function which defines our system
we turn to the functionals characterizing its dynamics,
the generator functional for connected Green functions and
its Legendre transform. We show finally that when Gauss' law,
i.e. the Coulomb interaction is properly established for the
external sources then the effective action simplifies
in what it includes a special combination of the external sources only.

\subsection{Partition function}
The partition function
\be
Z=\tr e^{-\beta (H-\mu)},
\ee
serves as a projection onto the ground state
as $\beta=1/T\to\infty$. The Hamiltonian $H=H_1+H_2$ is given by
\bea\label{hamo}
H_1&=&\sum_{s=1}^{n_s}\int_\mb{x}\psi^\dagger_{s,\mb{x}}
\left(-{\hbar^2\over2m}\Delta+\ve_\mb{x}\right)\psi_{s,\mb{x}},\nonu
H_2&=&\hf\sum_{s,s'=1}^{n_s}\int_{\mb{x},\mb{x'}}\psi^\dagger_{s,\mb{x}}
\psi^\dagger_{s',\mb{x'}}{e^2\over4\pi|\mb{x}-\mb{x'}|}
\psi_{s',\mb{x'}}\psi_{s,\mb{x}},
\eea
where $\int_\mb{x}=\int d^3x$ and we introduce the field variable
\be\label{sqop}
\psi_{s,\mb{x}}=\sum_{n=1}^\infty c_{s,n}\Psi_{n,\mb{x}},
\ee
with $\int_x\Psi^*_{n,x}\Psi_{m,x}=\delta_{n,m}$.
The coefficient $c_{s,n}$ is taken to be the destruction operator,
$c_{s,n}=a_{s,n}$, for a single-particle state of the wave function
$\Psi_{n,x}$ in the operator representation and becomes a Grassmann
variable in the path integral. The partition function is given in
the latter case as
\be
Z=\int\cd{\psi}\cd{\psid}\cd{u}e^{-S[\psid,\psi,u]},
\ee
with
\be\label{action}
S[\psid,\psi,u]=\int_x\left[\sum_s\psid_{s,x}\left(
\partial_t-{\hbar^2\over2m}\Delta-\mu+\ve_\mb{x}+ieu_x\right)
\psi_{s,x}+\hf(\nabla u_x)^2\right].
\ee
We use the notation $x=(\mb{x},t)$ and $\int_x=\int d^3 x\int_0^\beta dt$
 with the Euclidean time $t\in [0,\beta]$ and
the normalization $\int\cd{u}exp-\hf\int_x(\nabla u)^2=1$ is assumed
for the photon functional integral. An UV cut-off was introduced
to regulate the path integral and the normal ordering of $H_2$,
$\ve_\mb{x}\to\ve_\mb{x}+e^2/\Delta_{\mb{x},\mb{x}}$ was performed
in arriving at the action \eq{action}. 
Note that $\hbar$ is absent in the exponent of the
Euclidean path integral and its only role is to control  the
non-commutativity of the kinetic and the potential energies.
The Grassmann integral measure covers the
trajectories in the functional space spanned by the one-particle wave
functions
$\psi_{s,\mb{x},t}=\sum_nc_{s,n,t}\Psi_{n,\mb{x}}$ satisfying anti-symmetric
boundary conditions, $c_{s,n}(t+\beta)
=-c_{s,n}(t)$,
\be
\int\cd{\psi}=\prod_s\prod_n\int\cd{c_{s,n}}.
\ee
The range of $n$ extends over a complete set of
functions and the chemical potential $\mu$ is chosen in such a manner that
the ground state contains the desired number of particles.
For simplicity, we assume that all states are normalizable. When
de-localized states are needed the index $n$ has continuous
spectrum, as well. The time-component of the photon field
is purely imaginary for imaginary time as shown by the factor $i$
in the minimal coupling. The chemical potential, the homogeneous
component of the temporal component of the real time photon field
remains real.

Our basic assumption is that the formally $e$-independent
one-body interaction $H_1$ in Eq. \eq{hamo}
includes the singular part of the interactions and the two-body
interactions build up smoothly. We do not require the availability of the
perturbation expansion in $H_2$ instead we assume that the true ground
state, corresponding to the physical value of $e$, can be
achieved smoothly by increasing $e$ from zero in infinitesimal steps
without encountering singularities.

\subsection{Connected Green functions}
The source-dependent generator functional for the connected Green functions
is defined as
\bea\label{cggf}
e^{W[\sigma,j]}&=&\int\cd{\psi}\cd{\psid}\cd{u}e^{-S[\psid,\psi,u]
+i\sigma \cdot\sum_s\psid_s\psi_s+j\cdot u}\nonu
&=&\int\cd{u}e^{n_s\tr\log[G^{-1}+ieu-i\sigma]
-\hf\nabla u\cdot\nabla u+j\cdot u}\nonu
&=&\int\cd{u}e^{n_s\tr\log[G^{-1}+ieu]
-\hf[\nabla(u+\ei\sigma)]\cdot
[\nabla(u+{1\over e}\sigma)]+j\cdot(u+{1\over e}\sigma)}.
\eea
where the inverse propagator
\be\label{ivprop}
G^{-1}_{x,x'}=\delta_{x,x'}\left(\partial_t
-{\hbar^2\over2m}\Delta_\mb{x}-\mu+\ve_\mb{x}\right)
\ee
together with the notation $f\cdot g=\int_xf_xg_x$,
$\delta_{x,x'}=\delta_{\mb{x},\mb{x'}}\delta_{t,t'}$,
$\delta_{\mb{x},\mb{x'}}=\delta(\mb{x}-\mb{x'})$, and
$\delta_{t,t'}=\delta(t-t')$ was introduced.
Notice that the spatial gradient is diagonal in time,
$(\nabla)_{\mb{x},t,\mb{x'},t'}=\delta_{\mb{x},\mb{x'}}\delta_{t,t'}
\nabla_{\mb{x}}$.

It is instructive to compare the role the field
$\sigma$ and the
Grassmann-valued sources $\eta$, $\eta^\dagger$ play when the
action $S[\psid,\psi]\to S[\psid,\psi]-i\sigma\cdot\psid\psi
+\eta^\dagger\cdot\psi+\psid\cdot\eta$ is used for electrons
in the absence of the Coulomb interactions, $e=0$.
The dependence of the path integral on the sources $\eta$,
$\eta^\dagger$ is easy to find by diagonalising $H_1$. On the contrary,
the dependence on $\sigma$ cannot be obtained in a closed form.
The sources $\eta$, $\eta^\dagger$ control one-particle processes:
the introduction or removal of particles, without influencing their
states, the eigenvectors of $H_1$. The source $\sigma$
leaves the particle number unchanged but modifies the
states  they occupy, induces a polarization, and its
effect cannot be obtained in a simple form in terms of the
eigenvectors of $H_1$.

The imaginary time electron propagator
\be\label{itpdef}
G_{x,x'}=\la0|T\left[\psi_x\psid_{x'}\right]|0\ra
\ee
where $|0\ra$ denotes the ground state
is usually written in the spectral representation
\be\label{fgrsp}
G_{x,x'}=-\sum_{n=1}^\infty\int_\omega{\phi_{n,\mb{x}}\phi^*_{n,\mb{x'}}
\over-i\omega+E_n-\mu}e^{-i(t-t')\omega}
=-\sum_{n=1}^N\phi_{n,\mb{x}}\phi^*_{n,\mb{x'}}e^{-(t-t')(E_n-\mu)},
\ee
for $-\beta\le t-t'\le0$, $\beta\to\infty$, where
\be
 \left(-{\hbar^2\over2m}\Delta_\mb{x}+\ve_\mb{x}\right)\phi_{n,\mb{x}}
=E_n\phi_{n,\mb{x}},
\ee
and $\int_\omega=\int d\omega/2\pi$. The set of wave functions
$\phi_n$ and the spectrum $E_n$ correspond to non-interacting
quasi-particles in the presence of the external potential. We shall need
a parameterization of the quasi-particles dressed by the Coulomb
interaction. This will be achieved by writing the field operator \eq{sqop}
in terms of the dressed basis $\Psi_n$ specified later,
\bea\label{fgrspi}
G_{x,x'}&=&-\sum_{n,n'=1}^N\Psi_{n,\mb{x}}\Psi^*_{n',\mb{x'}}
\la0|a_ne^{-(H-\mu)(t-t')}a^\dagger_{n'}|0\ra\nonu
&=&\sum_{n=1}^Ne^{-(E_n-\mu)(t-t')}\Psi_{n,\mb{x}}\Psi^*_{n,\mb{x'}}
+\sum_{n,n'=1}^\infty g_{n,n'}(t-t')\Psi_{n,\mb{x}}\Psi^*_{n',\mb{x'}},
\eea
for $t>t'$ where the matrix elements of the time evolution operator are
parameterized as
\be
-\la0|a_ne^{-(H-\mu)t}a^\dagger_{n'}|0\ra=\delta_{n,n'}e^{-(E_n-\mu)t}
+g_{n,n'}(t).
\ee
The relation
\be\label{enhg}
\la0|a_n(H-\mu)a^\dagger_n|0\ra=E_n-\mu-\partial_tg_{n,n}(0)
\ee
together with the identification of $E_n$, $n>N$ with the energy increase
when a particle is added requires
\be\label{enh}
\partial_tg_{n,n}(0)=\cases{E_n-\mu&$n\le N$,\cr0&$n>N$.}
\ee
The part $n\le N$ of this equation holds due to $a^\dagger_n|0\ra=0$.
Notice that $g_{n,n'}(t)=\ord{t}$, when $\beta\to\infty$, assuming
that the ground state $|0\ra$ contains always $N$ particles, i.e.
the chemical potential is readjusted whenever the basis $\Psi_n$ is
modified.
Finally, the functional trace can be written as
\be
\tr A=\int_{\mb{x},t}A_{\mb{x},t,\mb{x},t}=\sum_{n=1}^\infty\int dtA_{n,t,n,t}
\ee
where $A_{n,t,n',t'}=\int_{\mb{x},\mb{x'}}\phi^*_{n,\mb{x}}
A_{\mb{x},t,\mb{x'},t'}\phi_{n',\mb{x'}}$.

It is well-known that photon vertex functions display
singularities when $p\to0$. This corresponds to a
qualitatively different dependence on the constant and the slowly
varying components of the external sources.
Notice that a given Fourier mode of the external sources,
$\sigma_{p,\omega}$, or $j_{p,\omega}$ shifts or couples to the
photon field $u_{p,\omega}$, respectively. Since the restoring force of this
photon field component to the vanishing equilibrium value is
$\ord{p^2}$ the response of the system for the slowly varying
external sources will be strong. In order to follow the
smooth building up of the Coulomb interactions, we shall
consider inhomogeneous external sources and their polarization
effects only, i.e. we set $\sigma_{p=0,\omega}=j_{p=0,\omega}=0$
in agreement with the absence of integration over the homogeneous,
$p=\omega=0$ modes of the photon field in QED \cite{qedhm}.

\subsection{Effective action}
The quantity of central importance is the effective action, the
functional Legendre transform of the generator functional
for connected Green functions. The reason this functional proves
to be useful for strongly coupled particles is that (i) its minimum
reproduces the free energy at finite temperature and the ground state
energy at zero temperature, and (ii) its value at
extended configurations gives the free energy or energy
corresponding to a given constraint, cf. Appendix \ref{effapex}, and
finally there are well
developed methods for its computation. The property (i) enables
us to compute the ground state energy in the low temperature limit.
The search for excited states is facilitated by feature (ii).

The effective action is defined as
\be\label{legendr}
\Gamma[\rho,v]=-W[\sigma,j]+i\sigma\cdot\rho+j\cdot v ,
\ee
where the induced charge density and photon field are given by
\be\label{lkv}
\rho_x={\delta W[\sigma,j]\over\delta i\sigma_x},~~~
v_x={\delta W[\sigma,j]\over\delta j_x}.
\ee
The inversion of these relations results in
\be\label{inv}
i\sigma_x={\delta\Gamma[\rho,v]\over\delta\rho_x},~~~
j_x={\delta\Gamma[\rho,v]\over\delta v_x}
\ee
showing that the effective action has an extremum, obviously
minimum in the ground state, $j=\sigma=0$.

The free energy for a given density configuration $\rho_x$
and external potential $v_x$, $T\Gamma[\rho,v]$,
is given by the functional integral
\be
e^{-\Gamma[\rho,v]}=\int\cd{u}e^{n_s\tr\log[G^{-1}+ieu]
+\hf(u+{1\over e}\sigma)\cdot\Delta\cdot
(u+{1\over e}\sigma)-i\sigma\cdot\rho+j\cdot(u+{1\over e}\sigma-v)},
\ee
where the sources are found by inverting Gauss' law, Eq. \eq{lkv},
\bea\label{forter}
ie\rho_x&=&e^{-W[\sigma,j]}\int\cd{u}\left[
\Delta\left(u_x+{1\over e}\sigma_x\right)+j_x\right]
e^{n_s\tr\log[G^{-1}+ieu]+\hf(u+{1\over e}\sigma)\cdot\Delta\cdot
(u+{1\over e}\sigma)+j\cdot(u+{1\over e}\sigma)}\nonu
&=&\Delta\left(\la u_x\ra_{\sigma,j}+{1\over e}\sigma_x\right)+j_x,\nonu
v_x&=&\la u_x\ra_{\sigma,j}+{1\over e}\sigma_x.
\eea

A particularly useful relation connecting the second functional derivatives
of these functionals is
\be\label{wg}
\int_{x'}\pmatrix{{\delta^2W[\sigma,j]\over
\delta i\sigma_x\delta i\sigma_{x'}}&
{\delta^2W[\sigma,j]\over\delta i\sigma_x\delta j_{x'}}\cr
{\delta^2W[\sigma,j]\over\delta j_x\delta i\sigma_{x'}}&
{\delta^2W[\sigma,j]\over\delta j_x\delta j_{x'}}}
\pmatrix{{\delta^2\Gamma[\rho,v]\over\delta\rho_{x'}\delta\rho_{x''}}&
{\delta^2\Gamma[\rho,v]\over\delta\rho_{x'}\delta v_{x''}}\cr
{\delta^2\Gamma[\rho,v]\over\delta v_{x'}\delta\rho_{x''}}&
{\delta^2\Gamma[\rho,v]\over\delta v_{x'}\delta v_{x''}}}
=\delta_{x,x''}\pmatrix{1&0\cr0&1},
\ee
which can be obtained by evaluating the trivial functional derivatives like
\bea
&& \delta_{x,x''}=  {\delta \rho_x\over \delta \rho_{x''}}
= {\delta \over \delta \rho_{x''}} {\delta W[\sigma,j]
\over \delta i\sigma_x}
=\int_{x'}\biggl[
 {\delta^2W[\sigma,j]\over \delta i\sigma_{x'}\delta i\sigma_x}
 {\delta i\sigma_{x'}\over \delta\rho_{x''}}
+ {\delta^2W[\sigma,j]\over \delta j_{x'}\delta i\sigma_x}
{\delta j_{x'}\over \delta\rho_{x''}}
 \biggr]\nonu
&&
=\int_{x'}\biggl[
 {\delta^2W[\sigma,j]\over \delta i\sigma_{x'}\delta i\sigma_x}
{\delta^2 \Gamma[\rho,v]\over \delta\rho_{x''}\delta\rho_{x'} }
+ {\delta^2W[\sigma,j]\over \delta j_{x'}\delta i\sigma_x}
{\delta^2 \Gamma[\rho,v]\over \delta\rho_{x''}\delta v_{x'} }
\biggr],
\eea
etc. according to the chain rule.

Expression \eq{cggf} for the generator functional for the
connected Green functions shows that the role of the test-potential
$v_x$ is played by the source $i\sigma$. The effective action
$\Gamma[\rho,v]$ gives the free energy of the system in the presence of
such external sources which generate the charge density and photon field
$\rho$ and $v$, respectively, in the ground state. The photon dynamics is not
traced by the density functional $\Gamma[\rho]$ and the Legendre
transformation yields the straightforward result
\be
E_{HK}[\rho]=\lim_{\beta\to0}{1\over\beta}\Gamma[\rho].
\ee
The change of variable $\sigma\to\rho$ is invertible
so long the Jacobian $\delta\rho/\delta\sigma$ is non-singular.
The latter is the connected two-point function which is positive definite
in a non-degenerate ground state. The density functional is thus
well-defined in a sufficiently small vicinity of the ground state density,
$\rho\approx\rt$. The only way the density functional may cease to exist
is the singularity of the Jacobian $\delta\sigma/\delta\rho$. According to
the identity \eq{wg}, this amounts to having a diverging eigenvalue for
the density-density correlation function, the indication of an instability
in the ground state.

\subsection{Gauge invariant sources}
The two external sources, $\sigma$ and $j$, introduced in the
generator functional \eq{cggf} are not independent physically.
In fact, the electrically charged environment of the system
can be represented either by its external charge distribution,
$\rho_{\mathrm{ext}}$, or by its external electric field,
$u_{\mathrm{ext}}$ which are related,
$\Delta u_{\mathrm{ext}}-e\rho_{\mathrm{ext}}=0$.
This is Gauss' law which arises from the redundancy, the
gauge invariance of the external sources. Since $\sigma/e$
and $j$ appear as external electric field or density,
$\Delta\sigma$ is identical with $ej$, in other words
one expects no dependence on $\Delta\sigma-ej$ in $W[\sigma,j]$.
In a more formal manner, the generator functional must be invariant
under the transformation
\be\label{dintr}
\sigma\to\sigma-e\chi,~~~j\to j+\Delta\chi.
\ee
Such an invariance arises from the fact that there are two independent
variables in Eq. \eq{cggf} but only one non-trivial functional,
$\tr\log[G^{-1}+ieu-i\sigma]$. The relation \eq{forter}
can be written as
\be\label{dsf}
{\delta W^\mr{tot}[\sigma,j]\over\delta\sigma}=\ei\Delta\cdot
{\delta W^\mr{tot}[\sigma,j]\over\delta j},
\ee
where $W^\mr{tot}[\sigma,j]=W[\sigma,j]+j\cdot{1\over2\Delta}\cdot j$
showing that $W^\mr{tot}[\sigma,j]$ actually depends
on the combination $J=\sigma+{e\over\Delta}\cdot j$ only.
The choice $\chi=\sigma/e$ and $\chi=-\Delta^{-1}\cdot j$ gives
\be\label{restw}
W[\sigma,j]=W\left[0,j+\ei\Delta\sigma\right]
+{1\over2e^2}\sigma\cdot\Delta\sigma+\ei\sigma\cdot j\nonu
=W\left[\sigma+{e\over\Delta}\cdot j,0\right]
-j\cdot{1\over2\Delta}\cdot j,
\ee
and the relation $W^\mr{tot}[J]=W[J,0]$. It is easy to find the physical
meaning of $W^\mr{tot}[J]$. In fact, according to gauge invariance
or Gauss' law the external source $j$ is coupled not only to
the photon field $u$ but has its contribution, $\Delta^{-1}\cdot j$
to the photon field itself. The gauge invariant partition function
of the electrons and external fields is
\bea
e^{W^\mr{tot}[J]}&=&\int\cd{u}e^{n_s\tr\log[G^{-1}+ieu-i\sigma]
+\hf(u+j\cdot\ide)\cdot\Delta(u+\ide\cdot j)}\nonu
&=&\int\cd{u}e^{n_s\tr\log[G^{-1}+ieu]+\hf(u+\ei J)\cdot\Delta(u+\ei J)}.
\eea

The reduction of the effective action to a functional with a single
variable can be achieved by rewriting Eq. \eq{forter} as
\be
{\delta\Gamma[\rho,v]\over\delta v}=ie\rho-\Delta v,
\ee
whose solution is
\be\label{ghrho}
\Gamma[\rho,v]=\Gamma[\rho,0]+ie\rho\cdot v-\hf v\cdot\Delta v.
\ee
The three terms on the right hand side are the free energy for a given
density and photon field as the sum of a mechanical term, the contribution
of the electron orbits $\Gamma[\rho,0]$, the interaction between the
charges and the photons $ie\rho\cdot v$ and the free energy of photons
$-v\cdot\Delta v/2$.
We can construct the effective action for $\rho$, the density functional,
by minimizing $\Gamma[\rho,v]$ in $v$. The minimum is at
$v_{\mathrm{gr}}[\rho]=ie\Delta^{-1}\cdot\rho$, where
\be\label{denftd}
\Gamma[\rho]=\Gamma[\rho,0]-{e^2\over2}\rho\cdot\ide\cdot\rho.
\ee
This relation indicates that $\Gamma[\rho]$ is the complete free energy,
that is the sum of the contribution of the electron structure,
$\Gamma[\rho,0]$, representing the mechanical, correlation energy and
the Coulomb energy of charge density $\rho$. The relation
\be
{\delta\Gamma[\rho]\over\delta\rho}
={\delta\Gamma[\rho,v]\over\delta\rho}_{\vert v=v_{\mathrm{gr}}[\rho]},
\ee
proves that the density functional is actually the
Legendre transform of $W[\sigma,0]$, or what will be useful
for the computations below, of the generator functional of the
complete electron-source system, $W^\mr{tot}[\sigma]$,
\be\label{dfltr}
-\Gamma[\rho]+i\sigma\cdot\rho=W[\sigma,0]=W^\mr{tot}[\sigma].
\ee
The reason to keep both sources $\sigma$ and $j$ and work with the  functionals
$W[\sigma,j]$ and its Legendre transform
\be\label{kvpev}
\Gamma[\rho,v]=\Gamma[\rho]+{e^2\over2}\rho\cdot\ide\cdot\rho
+ie\rho\cdot v-\hf v\cdot\Delta v
\ee
instead of using a simpler single-variable functional
will be explained in the next section.

\section{Non-interacting electrons}
There are two different problems intermingled in computing
the density functional $\Gamma[\rho]$. One is the expression of
the external source, the external potential of the density
functional theory, $\sigma$ in terms of the ground state density,
$\rho$ when the Legendre transformation is applied to  the non-interacting
electron problem. We shall explore this problem in this
section. We show in Appendix \ref{lnonint} how to obtain
the same result by solving a functional differential equation.
The other problem, left for the subsequent section,
is to take into account the Coulomb interaction.

We start with the generator functional for non-interacting electrons
\be\label{freefi}
W^\mb{free}[\sigma]=n_s\tr\log Q^{-1}
=n_s\tr\log G^{-1}-n_s\sum_{n=1}^\infty{1\over n}\tr(G\cdot i\sigma)^n
\ee
where the notation $Q_{x,y}^{-1}=\left(G^{-1}-i\sigma\right)_{x,y}$
was introduced. Here $\sigma$ is considered as a diagonal operator acting
on the space-time
$\sigma_{\mb{x},t,\mb{x'},t'}
=\delta_{\mb{x},\mb{x'}}\delta_{t,t'}\sigma_{\mb{x},t}$,
$(G\cdot\sigma)_{x,x'}=G_{x,x'}\sigma_x.$ It is easy to understand
the role of the infinite series in this equation.
The free energy of non-correlated particles in the presence of an
(Euclidean) external electric potential $i\sigma/e$ is $\rho\cdot i\sigma$,
where $\rho$ is the density. This is the term $n=1$ in Eq. \eq{freefi}.
We have to correct this expression if the particles are correlated at
different space-time points.
In order to take into account the influence say of a particle at point
$a$ on the probability of finding another particle at point $b$, we
add a term $i\sigma\cdot C^{(2)}\cdot i\sigma$ to the free energy
where $C^{(2)}_{a,b}$ is the connected density-density correlation
function. The contribution with a given $n$ in Eq. \eq{freefi}
corrects for the $n$-particle correlations in the free energy.

For the effective action we need the inversion of the equation
\be\label{vezsr}
\rho_x=-n_sQ_{x,x}=
-n_s\sum_{n=0}^\infty\left[(G\cdot i\sigma)^n\cdot G\right]_{x,x}.
\ee
According to Feynman rules, the external source $\sigma$ is
attached to graphs by the particle-hole propagator
\be\label{twoppr}
\tG_{x,y}=-n_s G_{x,y}G_{y,x}.
\ee
Therefore one expects that $\rho$, the Legendre-pair of $i\sigma$, will
appear in the graph in the combination $r=\tGi\cdot(\rho-\rt)$.
Since the charge density is non-vanishing in the ground state,
\be\label{vacrho}
\rt_x=-n_sG_{x,x}
\ee
we look for $i\sigma$ as a functional power series
in $\rho-\rt$. In this manner the inversion of Eq. \eq{vezsr} will be
sought in the form
\be\label{sinv}
i\sigma_a=\sum_{n=1}^\infty\int_{a_1,\ldots,a_n}{1\over n!}
A^\mb{free}_{a,a_1,\ldots,a_n}r_{a_1}\cdots r_{a_n} .
\ee
The index pairs $(x_a,t_a)$ are denoted by Latin indices $a$, etc.
By inserting this expansion and Eq. \eq{freefi} for
$Q$ into Eq. \eq{vezsr} and comparing
the terms with identical order in $\rho$ on both sides up to $\ord{r^3}$
we find
\be
A^\mb{free}_{a,b}=\delta_{a,b},~~~
A^\mb{free}_{a,b,c}=2n_s\int_d\tGi_{a,d} S_{d,b,c},~~~
A^\mb{free}_{a,b,c,d}=3!n_s \int_e \tGi_{a,e}\biggl[
S_{e,b,c,d} +2n_s\int_{f,g}S_{e,b,f}\tGi_{f,g}S_{g,c,d}\biggr]
\ee
with
\be
S_{a_1,a_2, \ldots,a_n}={1\over n!}\sum_{P\in S_n}
G_{a_{P(1)},a_{P(2)}}G_{a_{P(2)},a_{P(3)}} \cdots G_{a_{P(n)},a_{P(1)}}
\ee
where the sum extends over all permutations of $n$ indices.
These expressions for the free
gas correspond to the zeroth order ones in $e$ for the interacting gas.
In the last step we insert \eq{sinv} in the definition of $\Gamma$,
and rewrite the result in terms of the matrices
\be
\tS_{x_1,x_2,\ldots,x_n} = n_s\int_{a_1,a_2,\ldots,a_n}
S_{a_1,a_2,\ldots,a_n}\tGi_{a_1,x_1}\tGi_{a_2,x_2} \cdots \tGi_{a_n,x_n}
\ee
as
\be\label{gafree}
\Gamma^\mb{free}[\rho]=-W^\mb{free}[\sigma]+i\sigma\cdot\rho\nonu
=\Gamma^\mb{free}_0+\sum_{n=2}^\infty\int_{a_1,\ldots,a_n}{1\over n!}
\Gamma^\mb{free}_{a_1,\ldots,a_n}(\rho-\rt)_{a_1}\cdots(\rho-\rt)_{a_n}
\ee
with
\be\label{gaexfr}
\Gamma^\mb{free}_0= -n_s\tr\log G^{-1},~~
\Gamma^\mb{free}_{a,b}=\tGi_{a,b},~~
\Gamma^\mb{free}_{a,b,c}=2\tS_{a,b,c},~~
\Gamma^\mb{free}_{a,b,c,d}=3!\tS_{a,b,c,d}+2\cdot3!\int_{e,f}
\tS_{a,b,e}\tG_{e,f}\tS_{f,c,d}.
\ee
The interpretation of the form \eq{gafree} goes along the same
line as those of the free energy \eq{freefi}, namely
$\Gamma^\mb{free}_{a_1,\ldots,a_n}$ with $k\le n$ different space-time
point arguments represents the contribution to the free energy
due to the $k$-particle correlations when the particle density
is modified compared to the ground state.

At the minimum $\rho=\rt$ we find
the ground state energy
\be
E_{gr} =-\lim_{\beta\to\infty}{1\over\beta} n_s \tr\log G^{-1},
\ee
by assuming that the matrix $\tG$ is positive semidefinite.

\section{Evolution equation}
We tackle the second problem mentioned above, of the Coulomb
interaction by a method motivated by the renormalization group.
For this end we introduce a parameter $\lambda$ by modifying the action
$S\to S_\lambda$
\be
S_\lambda[\psid,\psi,u]=\int_x\left[\sum_s\psid_{s,x}\left(
\partial_t-{\hbar^2\over2m}\Delta-\mu+\ve_\mb{x}+ieu_x\right)\psi_{s,x}
+{1\over2\lambda^2}(\nabla u_x)^2\right].
\ee
For $\lambda<<1$ the Coulomb interaction is negligible
and the real physical problem is recovered at $\lambda=1$ because
$\lambda$ controls the photon field fluctuations. After the rescaling
$u\to\lambda u$ we find the action
\be\label{evaction}
S_\lambda[\psid,\psi,u]=\int_x\left[\sum_s\psid_{s,x}\left(
\partial_t-{\hbar^2\over2m}\Delta-\mu+\ve_\mb{x}+ie\lambda u_x\right)\psi_{s,x}
+\hf(\nabla u_x)^2\right],
\ee
which can be obtained from Eq. \eq{action} by the replacement
$e\to\lambda e$. Note that the source term contribution in the generator
functional, \eq{cggf} is not scaled for simplicity.

The functional differential equation,
\be\label{eveq}
\partial_\lambda\Gamma_\lambda[\rho,v]={\cal F}[\Gamma_\lambda[\rho,v];
\lambda,\rho,v]
\ee
obtained below for the effective action of the model with a given
control parameter value will then be integrated to find the true
effective action,
\be\label{flow}
\Gamma[\rho,v]=\Gamma_1[\rho,v]
=\Gamma_{\lambda_0}[\rho,v]+\int_{\lambda_0}^1d\lambda
{\cal F}[\Gamma_\lambda[\rho,v];\lambda,\rho,v].
\ee
For a sufficiently small initial value, $\lambda_0\approx0$, the initial
system is perturbative and $\Gamma_{\lambda_0}$ can be replaced
by the one-loop expression \eq{onelef}.

The evolution equation \eq{eveq} can be obtained by noting
\bea\label{wev}
\partial_\lambda\Gamma_\lambda[\rho,v]&=&-{d\over d\lambda}
W_\lambda[\sigma,j]\nonu
&=&iee^{-W_\lambda[\sigma,j]}\int\cd{\psi}\cd{\psid}\cd{u}
\int_x\psid_x\psi_xu_x
e^{-S_\lambda[\psid,\psi,u]+i\sigma\cdot\sum_s\psid_s\psi_s+j\cdot u}\nonu
&=&ie\int_x\la\psid_x\psi_xu_x\ra_{\sigma,j}\nonu
&=&ie\int_x\left(\la\psid_x\psi_xu_x\ra_{c\sigma,j}
+\la\psid_x\psi_x\ra_{\sigma,j}\la u_x\ra_{\sigma,j}\right),
\eea
where the subscript $c$ in the expectation value denotes the connected
part of the appropriate Green function. We express the connected Green
function by means of the functional derivative of the generator
functional $W[\sigma,j]$ and find
\be
\partial_\lambda\Gamma_\lambda[\rho,v]=ie\int_x
\left({\delta^2W_\lambda[\sigma,j]\over\delta i\sigma_x\delta j_x}
+{\delta W_\lambda[\sigma,j]\over\delta i\sigma_x}
{\delta W_\lambda[\sigma,j]\over\delta j_x}\right).
\ee
Eq. \eq{wg} is then used to arrive at a closed differential equation
\be
\partial_\lambda\Gamma_\lambda[\rho,v]=ie\left[\tr
\left({\delta^2\Gamma_\lambda[\rho,v]\over\delta\rho\delta v}\right)^{-1}
+i\rho\cdot v\right].
\ee

The reason to keep the dependence on the sources $\sigma$ and $j$ is
to arrive at the evolution equation \eq{wev} with the right hand side
that can be generated by the second functional derivative of the
generator functional $W$. But Eq. \eq{ghrho}
allows us to reduce the problem to functionals of a single variable.
The inverse of the operator
\be\pmatrix{{\delta^2\Gamma_\lambda[\rho,v]\over\delta\rho\delta\rho}&
{\delta^2\Gamma_\lambda[\rho,v]\over\delta\rho\delta v}\cr
{\delta^2\Gamma_\lambda[\rho,v]\over\delta v\delta\rho}&
{\delta^2\Gamma_\lambda[\rho,v]\over\delta v\delta v}}
=\pmatrix{{\delta^2\Gamma_\lambda[\rho,0]\over\delta\rho\delta\rho}
&i\lambda e\cr i\lambda e&-\Delta},
\ee
is
\bea
\pmatrix{{\delta^2W_\lambda[\sigma,j]\over
\delta i\sigma\delta i\sigma}&
{\delta^2W_\lambda[\sigma,j]\over\delta i\sigma\delta j}\cr
{\delta^2W_\lambda[\sigma,j]\over\delta j\delta i\sigma}&
{\delta^2W_\lambda[\sigma,j]\over\delta j\delta j}}
&=&\pmatrix{-\Delta\left(
-{\delta^2\Gamma_\lambda[\rho,0]\over\delta\rho\delta\rho}\Delta
+\lambda^2e^2\right)^{-1}
&-i\lambda e\left(-\Delta
{\delta^2\Gamma_\lambda[\rho,0]\over\delta\rho\delta\rho}
+\lambda^2e^2\right)^{-1}\cr
-i\lambda e\left(-{\delta^2\Gamma_\lambda[\rho,0]\over\delta\rho\delta\rho}
\Delta+\lambda^2e^2\right)^{-1}
&{\delta^2\Gamma_\lambda[\rho,0]\over\delta\rho\delta\rho}\left(
-\Delta{\delta^2\Gamma_\lambda[\rho,0]\over\delta\rho\delta\rho}
+\lambda^2e^2\right)^{-1}}\nonu
&=&\pmatrix{\left({\delta^2\Gamma_\lambda[\rho]
\over\delta\rho\delta\rho}\right)^{-1}
&i\lambda e\left({\delta^2\Gamma_\lambda[\rho]
\over\delta\rho\delta\rho}\right)^{-1}\cdot\ide\cr
i\lambda e\ide\cdot\left(
{\delta^2\Gamma_\lambda[\rho]\over\delta\rho\delta\rho}\right)^{-1}
&-\ide-\lambda^2e^2\ide\cdot\left(
{\delta^2\Gamma_\lambda[\rho]\over\delta\rho\delta\rho}\right)^{-1}
\cdot\ide}
\eea
according to Eqs. \eq{wg} and \eq{kvpev} and the
evolution equation can be written in the form
\be\label{seveq}
\partial_\lambda\Gamma_\lambda[\rho]=
-\lambda e^2\left\{\rho\cdot\ide\cdot\rho+\tr\left[
\left({\delta^2\Gamma_\lambda[\rho]\over\delta\rho\delta\rho}\right)^{-1}
\ide\right]\right\}
\ee
in terms of the density functional $\Gamma_\lambda[\rho]$. Note that this
is an exact equation, no small parameter was used in deriving it.
By writing the right hand side as
\be\label{evcoule}
\partial_\lambda\Gamma_\lambda[\rho]=\int_{\mb{x},\mb{y},t}
\left(\la\psid_{\mb{x},t}\psi_{\mb{x},t}
\psid_{\mb{y},t}\psi_{\mb{y},t}\ra_c+\rho_{\mb{x},t}\rho_{\mb{y},t}\right)
{\lambda e^2\over4\pi|\mb{x}-\mb{y}|}
=\lambda\alpha\int_{\mb{x},\mb{y},t}\la\psid_{\mb{x},t}\psi_{\mb{x},t}
\psid_{\mb{y},t}\psi_{\mb{y},t}\ra{1\over|\mb{x}-\mb{y}|}
\ee
where $\alpha=e^2/4\pi$,
we find that the evolution equation for $\Gamma[\rho]$ follows the
accumulation of the free energy of the electronic structure
in the absence of external photon field.
The evolution equation for $\Gamma[\rho,v]$
is more complicated because it takes into account an arbitrary,
externally prescribed photon field.

There are two kinds of contributions to the evolution equation:
First, the explicit $\lambda$-dependence in \eq{evaction} gives the resummation
of the perturbative contributions of the interactions between the
particles. Since we assumed
that the $e$-dependence of the ground state is continuous we
may have to adjust the chemical potential $\mu\to\mu_\lambda$
in order to keep the particle number $N_{tot}$ fixed in the
ground state. The chemical potential dependence is step function like
for $\beta\to\infty$ and the contribution to the evolution
equation containing $\partial_\lambda\mu_\lambda$ is made
vanishing by the adjustment. Another, implicit $\lambda$-dependence
is generated by the following optimization of the choice of quasi-particles.
The density functional will be minimized with respect to the
one-particle basis $\Psi_{n,\mb{x}}$ to take into account the interaction
with the external potential $\ve$. This variational step will be
carried out at each value of $\lambda$ and yields another contribution
to the evolution. This second $\lambda$-dependence which influences
mainly the ground state properties can be obtained by
performing the perturbation expansion of the HF equations in
$\Delta\lambda$ and retaining the leading order terms. In this manner
the usual iteration of the HF equation which is supposed to minimize
the (one-loop) energy functional will be carried out paralel
to the resummation of the loop expansion. The resulting
scheme maps the initial conditions, the non-interacting one-particle
wave functions into another set. This latter is used
to write the exact resolvent \eq{fgrsp} in the spectral
representation.

\section{Perturbative two-body forces}\label{ptbf}
Our goal in this section is to obtain the effective action
\eq{legendr} up to the terms quadratic in the fields
by first computing the generator functional \eq{cggf}
for the connected Green functions
and after then performing the Legendre transformation \eq{lkv}.
For this end we expand the fermion determinant of the third
line of Eq. \eq{cggf}.

We do not follow the perturbation expansion in $e$, instead
first we integrate out, at least formally, the electron field
and after then apply the saddle point expansion for the resulting path
integral over the photon field. On the one hand, the
improvement in following this strategy is that one can retain
an infinite subset of Feynman graphs arising from the
perturbation expansion in $e$. On the other hand, the non-systematic
nature of this scheme is that there is no small parameter
for the saddle point approximation of the photon path integral.
The remedy of this latter problem is to apply a cluster-expansion like
approximation scheme for the photon path integral, to truncate it
by ignoring the $n$-body correlations with a certain $n$.

First we recover the HF energy functional in the leading order,
$\ord{e^2}$ when the two-body correlations are retained. After that we point
out that the direct Coulomb and the exchange interactions correspond
to the leading order terms in the scheme where the two- and  three-body
correlations are ignored, respectively. Another point
hinting at the different dynamical origin of the direct and the
exchange terms is that they correspond to the tree and the
one-loop level evaluation of the photon path integral which indicates
that they describe correlations due to strong (classical)
electric field and due to correlated quantum fluctuations, respectively.

\subsection{Tree-level contribution}
Let us rewrite the generator functional for connected Green
functions given by  Eq. \eq{cggf} as
\be\label{efui}
e^{W[\sigma,j]}=\int\cd{u}e^{-S_\mr{eff}[u]}
\ee
by expanding the logarithmic function in the action $S_\mr{eff}[u]$
of the effective photon theory,
\bea
S_\mr{eff}[u]&=&-W^\mb{free}[-eu]
-\hf\left(u+\ei\sigma\right)\cdot\Delta\cdot\left(u+\ei\sigma\right)
-j\cdot\left(u+\ei\sigma\right)\nonu
&=&-n_s \tr\log G^{-1}
+n_s\tr\sum_{n=1}^\infty{(-1)^n\over n}\left(G\cdot ieu\right)^n
-\hf\left(u+\ei\sigma\right)\cdot\Delta\cdot\left(u+\ei\sigma\right)
-j\cdot\left(u+\ei\sigma\right).
\eea
We consider Eq. \eq{efui} as the
definition of an effective theory for the photon field in the presence
of the electron system. Similarly to the interpretation of
\eq{freefi}, the infinite series  in the effective action
corrects for the correlations between the electrons when propagating
on a given external background.

The loop expansions of the original functional integral
\eq{cggf} and of the effective theory \eq{efui} are not
equivalent because the trace in the exponent of \eq{efui}
comes from one-loop order of the Grassmann integral in \eq{cggf}.
Nevertheless one can apply the saddle point expansion to the
effective theory. On the tree-level of this scheme we approximate
the path integral by the maximum of the integrand,
$W[\sigma,j]\approx W^\mr{tree}[\sigma,j]$, where $W^\mr{tree}[\sigma,j]$
is the value of the effective action taken at its minimum
$u=u_\mr{cl}$. The saddle point $u_\mr{cl}$ satisfies the equation
\be\label{eqmo}
{\delta S_\mr{eff}[u_\mr{cl}]\over\delta u_x}=-ien_s\sum_{n=0}^\infty(-1)^n
\left[\left(G\cdot ieu_\mr{cl}\right)^nG\right]_{x,x}
-\left[\Delta\cdot(u_\mr{cl}+\ei\sigma)\right]_{x}-j_x=0 .
\ee
When the terms $\ord{(eu_\mr{cl})^3}$
in the exponent of \eq{efui} beyond the polarization of the ground
state are neglected then we find
\be\label{ucl}
u_\mr{cl}=D\cdot\left(j-ie\rt+\ei\Delta\cdot\sigma\right),
\ee
where
\be\label{phopr}
D^{-1}=-\Delta+e^2\tG
\ee
stands for the photon propagator.
We can write the corresponding electric energy
by introducing the induced electric field in the ground state,
$\mr{E}=\nabla u_\mr{cl}$, as
\bea\label{trece}
W^\mr{tree}[\sigma,j]&=&-S_\mr{eff}[0]+\hf
{\delta S_\mr{eff}[0]\over\delta u}\cdot
\left({\delta^2S_\mr{eff}[0]\over\delta u\delta u}\right)^{-1}
\cdot{\delta S_\mr{eff}[0]\over\delta u}\nonu
&=&n_s\tr\log G^{-1}+{1\over 2e^2}\sigma\cdot \Delta\cdot\sigma
+\ei \sigma\cdot j+\hf\left(j-ie\rt+\ei\sigma\cdot\Delta\right)
\cdot D\cdot\left(j-ie\rt+\ei\Delta\cdot\sigma\right)\nonu
&=&n_s\tr\log G^{-1}
+\hf\int_x\mr{E}^2_x+{e^2\over2}u_\mr{cl}\cdot\tG\cdot u_\mr{cl}
+{1\over 2e^2}\sigma\cdot \Delta\cdot\sigma
+\ei \sigma\cdot j
+\ord{e^4}.
\eea
Here the first term in the last line is the non-interacting
contribution to the mechanical energy of the electron system,
the second and third terms represent the energy of the photon
field $u_\mr{cl}$ and of the polarization, respectively, whereas the last
two terms account for the energy of the external sources.

\subsection{HF energy}
The next step is to consider the path integral \eq{efui} in the one-loop
approximation and retaining the two-particle correlations only.
First we take the case $\sigma=0$, when we find
\be\label{wfj}
W[0,j]=W^\mr{tree}[0,j]+W^\mr{1-loop}=\hf(j-ie\rt)\cdot D\cdot(j-ie\rt)+W_0
\ee
where $W_0=n_s \tr\log G^{-1}+W^\mr{1-loop}$ and  the one-loop correction
$W^\mr{1-loop}$ is  independent of the source $j$,
\be\label{wnulla}
W^\mr{1-loop}=-\hf\tr\log{\delta^2S_\mr{eff}[0]\over\delta u\delta u}
=-\hf\tr\log D^{-1}+\hf\tr\log-\Delta.
\ee
We shall use the parameterization
\bea\label{phselfe}
\Sigma_{x,t,x',t'}^{ph}&=&-n_se^2
\Biggl[\sum_{n,m=1}^Ne^{(E_m-E_n)(t-t')}
\Psi_{n,\mb{x}}\Psi^*_{n,\mb{x}'}\Psi_{m,\mb{x}'}\Psi^*_{m,\mb{x}}\nonu
&&+\sum_{m=1}^N\sum_{n,n'}g_{n,n'}(t-t')e^{E_m(t-t')}
\Psi_{m,\mb{x}'}\Psi^*_{m,\mb{x}}\Psi_{n,\mb{x}}\Psi^*_{n',\mb{x}'}\nonu
&&+\sum_{n=1}^N\sum_{m,m'}g_{m,m'}(t-t')e^{-E_n(t-t')}
\Psi_{m,\mb{x}'}\Psi^*_{m',\mb{x}}\Psi_{n,\mb{x}}\Psi^*_{n,\mb{x}'}\nonu
&&+\sum_{n,n',m,m'}g_{n,n'}(t-t')g_{m,m'}(t-t')
\Psi_{n,\mb{x}}\Psi^*_{n',\mb{x}'}\Psi_{m,\mb{x}'}\Psi^*_{m',\mb{x}}\Biggr],
\eea
for the photon self-energy due to the particle-hole polarizations
$\Sigma^{ph}=-n_se^2\tG$ for $-\beta\le t-t'\le0$.
The complete generator functional is
\be\label{oneloop}
W[\sigma,j]=-\hf\sigma\cdot(\tG-e^2\tG\cdot D\cdot \tG)\cdot\sigma
+ie\sigma\cdot\rt+\ei j\cdot \sigma
+\hf(j-ie\rt)\cdot D\cdot(j-ie\rt)+W_0
\ee
according to Eq. \eq{restw}.

The leading order, $\ord{e^2}$ HF energy of the ground state
corresponds to $\sigma=j=0$ as $\beta\to\infty$,
\be\label{kotale}
E_{\mathrm{gr}}=-\lim_{\beta\to\infty}{1\over\beta}W[0,0]
=\lim_{\beta\to\infty}{1\over\beta}\Biggl\{
-n_s\tr\log G^{-1}-{e^2\over2}\rt\cdot {\ide}\cdot\rt
+{e^2\over 2}\tr\ide\cdot\tG+\ord{e^4}\Biggr\},
\ee
where the last term comes from
\be\label{phdee}
-\hf\tr\log-D^{-1}\cdot\ide=\hf\sum_{n=1}^\infty{1\over n}
\left(e^2\tG\cdot\ide\right)^n.
\ee
The first two terms in Eq. \eq{kotale} stand for the expectation value
of the one-body Hamiltonian $H_1$ in the state where $n_s$
particles are found with wave function $\Psi_{n,x}$, $n=1,\cdots,N$.
The third term represents the exchange energy,
\bea
-\hf e^2\int_{x,t,x',t'}\delta_{x,x'}\delta_{t,t'}
\left(-{1\over\Delta_x}\right)\tG_{x,t,x',t'}.
\eea
Thus, one obtains
\bea\label{hfee}
E_{\mathrm{gr}}&=&n_s\sum_{n=1}^N\int_\mb{x}
\Psi_{n,\mb{x}}^*\left[-{\hbar^2\over2m}\Delta+\ve_\mb{x}\right]\Psi_{n,\mb{x}}
+{n_s^2e^2\over2}\sum_{n,m=1}^N\int_{\mb{x},\mb{x'}}\Psi_{n,\mb{x}}
\Psi_{n,\mb{x}}^*
{1\over4\pi|\mb{x}-\mb{x}'|}\Psi_{m,\mb{x}'}\Psi_{m,\mb{x}'}^*\nonu
&&-{n_se^2\over2}\sum_{n,m=1}^N\int_{\mb{x},\mb{x}'}\Psi_{n,\mb{x}}
\Psi_{m,\mb{x}}^*
{1\over4\pi|\mb{x}-\mb{x}'|}\Psi_{m,\mb{x}'}\Psi_{n,\mb{x}'}^*+\ord{e^4}.
\eea
The different powers of $n_s$ in front of the direct and the
exchange terms remind us of their different origins. The
multiplicative factors $\ord{n_s^2}$ and $\ord{n_s}$
characterize the contributions of the fermionic and bosonic
determinants, respectively.

\subsection{Density functional}
We re-derive the HF equations in this section
by following the Legendre transformation \eq{legendr}
and minimizing the resulting energy density.
According to \eq{restw}, the
density functional satisfies the equation
\be\label{gawei}
-\Gamma[\rho]+i\sigma\cdot\rho=W\left[0,\ei\Delta\sigma\right]
+{1\over2e^2}\sigma\cdot\Delta\sigma.
\ee
Using the one-loop result, Eq. \eq{wfj}, the source $\sigma$
is to be eliminated in favor of the density $\rho$ by means of the
equation
\be
i\rho=\Delta\cdot D\cdot\left(\eik\Delta\sigma-i\rt\right)
+\eik\Delta\sigma+\ord{e^2}+\ord{\sigma^2}.
\ee
One has to be careful in determining the orders which are
reliable because of the factor $1/e$ appearing on the right
hand side of Eq. \eq{gawei} in the argument of $W$.
Due to this factor the higher order contributions in $\sigma$
will not be suppressed by $e$. As a result we have a double
expansion in $e$ and $\sigma$. By inverting these relations
one should use the perturbative expression
\be
D=-\ide-e^2\ide\cdot\tG\cdot\ide+\ord{e^4}
\ee
for the photon propagator as late in the computation as possible.
Instead, by relying on \eq{phopr} we find
\be
i\sigma=\tGi\cdot(\rho-\rt)+\ord{e^2}+\ord{\rho^2}
\ee
and
\be\label{densfm}
\Gamma[\rho]=\hf\rho\cdot\tGi\cdot\rho+\hf\rt\cdot\tcGi\cdot\rt
-\rho\cdot\tcGi\rt-W_0+\ord{\sigma^2}+\ord{e^4}
\ee
where the one-loop electron-hole inverse propagator $\tcGi$ has been
introduced,
\be\label{cgdef}
\tcGi=\tGi+{e^2\over\Delta},~~~
\tcG=\tG-\tG\cdot{e^2\over\Delta}\cdot\tG+\ord{e^4}.
\ee
It is instructive to compute the effective action without the
photon contributions,
\be
\Gamma[\rho,0]=\hf(\rho-\rt)\cdot\left(\tGi+{e^2\over\Delta}\right)
\cdot(\rho-\rt)-W_0+\ord{\sigma^2}+\ord{e^4},
\ee
showing that the density fluctuations around $\rt$ generate
two kinds of energy increases in the electron system: the
contributions corresponding to the terms $\tGi$ and $e^2/\Delta$ are
related to the mechanical correlations and Coulomb interactions,
respectively.

The complete effective action is found from Eq. \eq{kvpev},
\be\label{onelef}
\Gamma[\rho,v]=-n_s\tr\log G^{-1}
+\hf(\rho-\rt)\cdot\tcGi\cdot(\rho-\rt)
+ie\rho\cdot v-\hf v\cdot\Delta v+\hf\tr\log D^{-1}-\hf\tr\log-\Delta.
\ee
The first term is the sum of the single particle energies
at $\rho=\rt$, the second term represents the change of the
polarization energy due to the excitation
of the electron system, the third
one describes the interaction of the charge density with the electric
field, the fourth one is the energy of the electric field, and
before the last term contains the exchange energy.

The ground state energy functional is found in the low temperature limit as
\be\label{densf}
E_{\mathrm{gr}}[\rho,v]=\lim_{\beta\to\infty}{1\over\beta}\Gamma[\rho,v].
\ee
The minimum of the density functional is reached at
\be\label{onegr}
\rgr=\left(1+e^2\tG\cdot\ide\right)\cdot\rt,~~~
\vgr=ie\ide\cdot\rgr.
\ee
Inserting these in (\ref{densf}), one obtains the ground state energy \eq{hfee}
in the leading order. Hereby,  the first term on the r.h.s. of
\eq{onelef}, the fermion determinant provides the sum of the
single fermion energies,
the second term  does not
contribute at $\ord{e^2}$, the third and fourth terms give rise the
Coulomb-energy, whereas the exchange energy (fifth term) occurs due to
the boson determinant.
The charge density of the ground state is modified with respect to
the value $\rt$, but this does not influence the energy of the
ground state in the order $\ord{e^2}$.

The second functional derivatives of the effective action give the
inverse propagators. One might wonder why do the leading order inverse
propagators appear in the one-loop level effective action \eq{onelef}.
The inverse of the free photon propagator, $-\Delta$, for example
does not contain the polarization of the electron ground state.
But the latter can be taken into account by minimizing in the
density for a given external photon field configuration. In fact,
for a given $v$ the minimum of \eq{onelef} is reached at
$\rho_{\mathrm{gr}}[v]=-ie\tG\cdot v+\rt$ and the effective action for
the photon field only,
\be
\Gamma[\rho_{\mathrm{gr}}[v],v]
=-n_s\tr\log G^{-1}+\hf v\cdot(-\Delta+e^2\tG)\cdot v+ie\rt\cdot v
+\hf\tr\log D^{-1}-\hf\tr\log-\Delta
\ee
reproduces the one-loop inverse propagator. This is the reason
of the absence of the photon self-energy in $1/\Delta$
on the right hand side of the evolution equation \eq{seveq}.
Similar consideration gives
$\Gamma[\rho,v_{\mathrm{gr}}[\rho]]=\Gamma [\rho]$.

\subsection{Exchange-correlation energy}\label{exchi}
It is worthwhile noting the elusive source of the exchange
interaction. In the usual, variational
derivation of the HF equations the direct Coulomb and the exchange
terms correspond to the diagonal and the cross terms in
computing the expectation value of the Coulomb potential energy
operator sandwiched between Slater determinants. In the second quantized
formalism the exchange term comes neither from the
fermion determinant which incorporates the single-electron
aspects of the dynamics nor from the Slater determinant
because the states in the Fock-space are properly
(anti)symmetrized by construction, without using any determinant
based on non-symmetrical wave functions.
Instead it arises from the photon functional determinant,
showing its genuine quantum nature.

According to the remarks made after Eq. \eq{freefi},
the fluctuating photon field $u$ induces free energy changes
$\ord{u^n}$ through the $n$-particle density correlations,
the term corresponding to $n$ of the exponent in Eq. \eq{efui}.
On the tree level \eq{trece} we find the interplay of
the $\ord{u}$, local and the $\ord{u^2}$ two-electron correlation
contributions. They make up the direct Coulomb contribution
of the external sources and their polarization effects.
The common feature of graphs appearing
in the tree-level solution of the effective theory is
that the external legs end in the combination
$j-ie\rt-\Delta\sigma/e$, the inhomogeneous term in the
equation of motion, \eq{eqmo}. The external legs are attached
to the rest of the graph by the photon propagator, $D$. These
contributions describe the Coulomb energy of the local charge
distribution and the polarization induced by the external
local classical sources, $\sigma$ and $j$ and internal local
fluctuations, $\rt$.

The one-loop contribution, $W^\mr{1-loop}$,
is the 'zero point fluctuation' of the photon field.
The normal modes of the photon field in \eq{efui} are coupled
to the $\ord{u^2}$ two-particle correlators and $W^\mr{1-loop}$ is the
sum of these contributions to the energy.
The $n$-th order contribution of this
correlation in the perturbation expansion corresponds to the
term $n$ in the expansion \eq{phdee} where according to Wick's theorem
the $2n$ insertions of the photon field are broken up into
$n$ photon propagators connecting the density correlation functions.
The common element of the
one-loop level graphs is that they have no external legs since
$\sigma$, $j$ and $\rt$ couple to the first power of the photon field
(minimal coupling) and they drop out from the second functional
derivative of the effective action. These contributions
represent the Coulomb energy of the bi-local or higher order correlated
fluctuations, the leading order contribution being the bi-local
'exchange densities' introduced in Ref. \cite{slater}.

We considered above local densities and two-particle correlations.
When higher order correlations are retained in the equation of motion
\eq{eqmo} to determine the saddle point then higher order density
correlations appear on the tree and the one-loop level, as well.

The traditional direct Coulomb and exchange terms are the
leading order $\ord{e^2}$ contributions of the tree
and the one-loop level graphs, respectively. These and other,
higher order graphs of the tree and the one-loop sectors
can be related to each other by 'opening' the electron
lines of $\rt$ at the leg of a tree-level graph and attaching to
another electron line, obtained in a similar manner.
This explains that despite their difference in the
effects they incorporate these graphs happen to be
in the same order of $e$ and $\hbar$ (when restored for real time).
This may happen because the effective action \eq{efui} contains
the sum of the tree and one-loop level contributions of the
Grassmann integration and there is no further small parameter
for the saddle point expansion within the effective theory.

The exchange, maybe better to say correlation interaction
nevertheless makes its appearance
even on the tree-level of the effective theory, i.e. in the
classical approximation for the photon field.
In fact, the term $\ord{e^2}$ in the second
equation of \eq{trece} can be interpreted as the
contribution of the bi-local exchange density to the electric energy.

\section{Multi-local approximations}
We shall use in this section the evolution equation to perform
a partial resummation of the perturbation expansion in the
Coulomb interaction. The non-perturbative external potential is
taken into account variationally by minimizing the
effective action against the choice of the one-particle
wave functions $\Psi_{n,x}$ at each stage of the evolution.
The evolution equation will be studied in different truncation
schemes. For systems with homogeneous ground state the expansion
in the inhomogeneity of the excitation, the gradient expansion,
is a generally employed truncation scheme of the effective action.
The localized bound states render this method useless and
we have to turn to another classification scheme, alluded to in the
previous section, the cluster expansion.

Each truncation scheme $S$ will be characterized by a
subspace ${\cal A}_S$ of the space ${\cal A}$ of effective action
functionals and the evolution equation is approximated by projecting
it into ${\cal A}_S$. In order to make this
projection easy to define, the subspace ${\cal A}_S$ must be linear.
Terms like $\int_{x_1,\cdots,x_\ell}\gamma_{x_1,\cdots,x_\ell}
\rho_{x_1}^{n_1}\cdots\rho_{x_\ell}^{n_\ell}$
in the effective action represent the interactions between
$\ell$ clusters and will be called $\ell$-local.
A functional will be called $\ell$-local in general
if it contains $\ell$-local
clusters only. The linear subspace made up by the sum of $n$-local
functionals with $n\le\ell$ will be denoted by ${\cal L}_\ell$.
The effective action, $\Gamma[\rho]$,
contains multi-local terms with arbitrary high values of $\ell$.
The approximation $f_\ell$ where terms up to $\ell$-locality are
retained in a free, unconstrained manner corresponds to projecting
the effective action onto the subspace ${\cal A}_{f_\ell}={\cal L}_\ell$.

It will be important to preserve certain functional structures
within a given locality subspace. For this end we
refine our scheme and introduce the constrained $\ell$-local
approximation, $c_\ell$, based on the
subspace ${\cal A}_{c_\ell}$ of $\ell$-local functionals.
Some complications arise form the fact that the functionals
obtained by imposing a particular parameterization
generally do not span a linear space and the restriction of the
evolution equation for such a space is not uniquely defined.

The linear subspace of the mixed scheme $f_kc_\ell$
with $k<\ell$ will be constructed by considering
the functionals which are the sum of two functionals.
The first one is a constrained one ($c_\ell$).
The other belongs to the unconstrained class ($f_k$) but
chosen in such a manner that it does not interfere
with the previous component. This will be achieved
by orthogonalizing the two components. In such a manner
the subspace of the mixed scheme is defined as
${\cal A}_{f_kc_\ell}={\cal A}_{c_\ell}
+({\cal A}_{f_k}\cap{\cal A}_{c_\ell}^\bot)$
where ${\cal A}_{c_\ell}^\bot$ denotes the orthogonal
complement of ${\cal A}_{c_\ell}$. We want to avoid
the detailed definition of the functional spaces in question,
it is sufficient to say at this point that ${\cal L}_\ell$
is spanned by functionals of the form
$\int_{x_1,\cdots,x_k}\gamma_{x_1,\cdots,x_k}
\rho_{x_1}^{n_1}\cdots\rho_{x_\ell}^{n_k}$ and the scalar product
is chosen in such a manner that the functionals
$\rho_x^n$ and $\rho_x^m$ be orthogonal for $n\not=m$.

Once the subspace of the approximation has been identified,
both sides of the evolution equation
\eq{eveq} is projected onto ${\cal A}_{f_\ell}$ or
${\cal A}_{f_k}\cap{\cal A}_{c_\ell}^\bot$
in the scheme $f_\ell$ or $f_kc_\ell$, respectively. The evolution
of the constrained component in the latter case is determined
by imposing the evolution equation at certain suitably chosen
configurations $\rho_x$.

One may distinguish three levels of the description of the ground state of
 $N$ particles.
\begin{enumerate}
\item The ground state is characterised by a wave function of $3N$ variables.
\item The  spectral representation is evoked to parametrize the Green functions,
it is based on quasi-particle wave functions, $N$ of them corresponding to
the filled states and the rest for the virtual excitations.
\item A single function, the density is used and the coefficient functions
of the constrained scheme are treated on a space-time lattice without any
reference to wave functions.
\end{enumerate}
The first level is for the first quantized formalism, the second exploites
the simplifications inherent in second quantization. These levels are certainly
justified for small systems but lead to complications when the ground
state of a large number of particles is sought. In fact, the computational
time is at least $\ord{N^4}$ for a $c_2$ scheme at description level 2
(Hartree-Fock) and
definitely worse for level 1. Level 3 offers more effective algorithms
for large systems. Namely, after the diagonalization of a non-interacting
Hamitonian for the initial condition, a $\ord{N^3}$ step, the Coulomb interaction
can be taken into account in an $f_1c_2$ scheme. This latter is based
on the density Green function, a space-time matrix of dimension $N^p$.
The power $0\le p\le1$ depends on the resolution in the space-time
needed to reach the desired accuracy. In fact, for a fixed, particle number
independent resolution we have $p=0$ and one finds $p=1$ when the dimension of the
one-particle sector of the Fock space is kept proportional to the
particle number. The computational time is therefore $\ord{N^{2p}}$,
slower increasing with $N$ than for the diagonalisation of the non-interacting
Hamiltonian in the initial condition.

An $f_1c_2$ scheme of level 3 is presented first in Section \ref{matrix},
followed by a number of approximations of level 2.
A simple $c_2$ scheme where we neglect the terms $\ord{e^4}$
and restrict ourselves to the parametrized bi-local
approximation reproduces the HF equations.
A possible simple improvement is an $f_1c_2$ scheme
which may go beyond the leading order of the perturbation expansion.
The most involved scheme sketched in this paper is $f_2$.
Finally we mention a modified $f_1c_2$ approximation leading
to a generalization of the KS scheme.

\subsection{Quasi-particles}
The internal structure of an extended state is characterized
in perturbative Quantum Field Theory by form factors. In the
functional setting quadratic functionals usually
correspond to free point-like particles and the form factors
are hidden in the non-quadratic, higher order terms.
This situation can be changed by considering the propagator,
the kernel of the quadratic part as a free variable \cite{vasil},
occuring on the description level 3.We choose here a simpler short-cut
for the level 2 description by parameterizing
the quadratic kernel by means of one-particle wave functions,
as given in Eq. \eq{fgrsp}. This expression will serve as the definition of
the quasi-particles making up the ground state and the
excitations above it.

The success of the effective action as a quadratic functional in the
density suggests the form
\be\label{genan}
\Gamma[\rho]=\hf(\rho-\rt)\cdot\tcGi\cdot(\rho-\rt)
-{\lambda^2e^2\over2}\rho\cdot\ide\cdot\rho+\cC[\rho]
\ee
where the `potential energy' $\cC[\rho]$ is introduced to take
into account the higher order radiative corrections, $\tcG$
and $\rt$ are given by \eq{cgdef} and \eq{vacrho}, respectively.
We write the evolution equation imposed for a fixed,
$\lambda$-independent $\rho$ as
\bea\label{evolph}
&&\hf(\rho-\rt)\cdot\partial_\lambda\tcGi\cdot(\rho-\rt)
-\partial_\lambda\rt\cdot\tcGi\cdot(\rho-\rt)+\partial_\lambda\cC[\rho]
=-\lambda e^2\sum_{n=0}^\infty(-1)^n\tr\left[\tcG\cdot\left(
\cC^{(2)}\cdot\tcG\right)^n\cdot\ide\right],
\eea
where the notation
\be
{\delta^2\Gamma\over\delta\rho_a\delta\rho_b}=\tcGi_{a,b}+\cC^{(2)}_{a,b}.
\ee
is introduced. Note that the summation on the right hand side of
\eq{evolph} can be constrained to $n\le\ell$ when the effective action is
sought in $\ell$-local form because $\cC^{(2)}$ contains ultra-local terms
in the density.
The evolution equation \eq{evolph} will be considered
in two different manners. (i) The evolution of the effective action at its
minimum, $\rho=\rho_\mr{gr}$, determines mainly the ground state
properties, i.e. the quasiparticle wave functions $\Psi_n$ and spectrum
$E_n$. We shall be looking for the generalization of the
HF and the KS schemes in this way.
(ii) The evolution of the functional dependence away from the minimum
determines the different potentials parameterizing
the increase of the effective action for $\rho\not=\rho_\mr{gr}$.
We shall not assume that the `excitation' $\rho-\rho_\mr{gr}$ is small.

(i) The minimum of the density functional is reached at
\be\label{mindf}
\rho_\mr{gr}={1\over1-\tcG\cdot{\lambda^2e^2\over\Delta}}
\left(\rt-\tcG\cdot{\delta\cC[\rho_\mr{gr}]\over\delta\rho}\right)
=\left(1+\tG\cdot{\lambda^2e^2\over\Delta}\right)\cdot
\left(\rt-\tcG\cdot{\delta\cC[\rho_\mr{gr}]\over\delta\rho}\right).
\ee
Making use of the definition \eq{cgdef} and the relation
\bea
\partial_\lambda\left[\Gamma[\rho]+{\lambda^2e^2\over2}\rho
\cdot\ide\cdot\rho\right]_{\vert\rho=\rho_\mr{gr}}
&=&\partial_\lambda\cC[\rho_\mr{gr}]
-\partial_\lambda\rt\cdot\left(1+{\lambda^2e^2\over\Delta}\cdot\tG\right)
\cdot\left({\lambda^2e^2\over\Delta}\cdot\rt
-{\delta\cC[\rho_\mr{gr}]\over\delta\rho}\right)\nonu
&&+\hf\left({\delta\cC[\rho_\mr{gr}]\over\delta\rho}
-\rt\cdot{\lambda^2e^2\over\Delta}\right)\cdot
\left(\tG\cdot{2\lambda e^2\over\Delta}\cdot\tG
-\partial_\lambda\tG\right)\cdot
\left({\delta\cC[\rho_\mr{gr}]\over\delta\rho}
-{\lambda^2e^2\over\Delta}\cdot\rt\right)
\eea
with $\partial_\lambda \cC [\rho_\mr{gr}]=
\partial_\lambda \cC[\rho]_{\vert{\rho=\rho_\mr{gr}}}$,
we find at the minimum
\bea\label{atmin}
&&-\partial_\lambda\rt\cdot{\lambda^2e^2\over\Delta}\cdot\rt
+\lambda e^2\tr\tcG\cdot\ide
=-\lambda e^2\sum_{n=1}^\infty(-1)^n\tr\left[\tcG\cdot\left(
\cC^{(2)}[\rho_\mr{gr}]\cdot\tcG\right)^n\cdot\ide\right]\nonu
&&-\hf\left({\delta\cC[\rho_\mr{gr}]\over\delta\rho}
-\rt\cdot{\lambda^2e^2\over\Delta}\right)\cdot
\left(\tG\cdot{2\lambda e^2\over\Delta}\cdot\tG
-\partial_\lambda\tG\right)\cdot
\left({\delta\cC[\rho_\mr{gr}]\over\delta\rho}
-{\lambda^2e^2\over\Delta}\cdot\rt\right)\nonu
&&+\partial_\lambda\rt\cdot{\lambda^2e^2\over\Delta}\cdot\tG
\cdot{\lambda^2e^2\over\Delta}\cdot\rt
-\partial_\lambda\rt\cdot\left(1+{\lambda^2e^2\over\Delta}\cdot\tG\right)
\cdot{\delta\cC[\rho_\mr{gr}]\over\delta\rho}
-\partial_\lambda \cC[\rho_\mr{gr}].
\eea
This equation is in general inconsistent with the evolution equation
imposed at $\rho\not=\rho_\mr{gr}$ since the functions parameterizing
the functional, the propagators, are overdetermined.
By following the strategy of turning the renormalization group
equations into a partial resummation
of the perturbation expansion we retain the modification of
our effective action within the framework of the ansatz
in the region where the approximation is supposed to be the best,
in the ground state $\rho=\rho_\mr{gr}$.

(ii) We shall use in this section the parameterization
$\cC=\cC_\mr{qfl}+\cC_\mr{exct}$ up to bi-local terms. The first term, chosen
originally as
\be
-n_s\tr\log G^{-1}+\hf\tr\log\cD^{-1}-\hf\tr\log-\Delta
\ee
and controlling mainly the zero point quantum fluctuations,
will be replaced by the simplified form
\be\label{cqfl}
\cC_\mr{qfl}[\rho]=-n_s\tr\log G^{-1}+\hf\tr\log\left[-D^{-1}\ide\right]
+\hf\tr\ D\cdot\sigma-{1\over4}\tr\ \sigma\cdot D\cdot\sigma\cdot D
\ee
according to the assumption about bilocality. The photon propagator
$\cal D$ is constructed by allowing the photon self-energy to
become a local function of the density,
\be\label{phoprsk}
\cD^{-1}_{a,b}=D^{-1}_{a,b}+\sigma(a,\rho_a)\delta_{a,b}
=[-\Delta+\sigma(a,\rho_a)]\delta_{a,b}+\Sigma^{ph}_{a,b}
\ee
as assumed in the KS scheme \cite{ksh}.
The explicit dependence of the self-energy on the position is due to the
localized single-particle states in the ground state and can be
omitted in the homogeneous electron gas. The second term of $\cC$
is supposed to govern the excitations
and will be the sum of an ultra-local and a bi-local contribution,
\be\label{vans}
\cC_\mr{exct}[\rho]=\int_xU(x,\rho_x)
+\sum_{n,m=0}^\infty{1\over2n!m!}\rho^n\cdot\gamma^{(n,m)}\cdot\rho^m
\ee
We carry out the replacement
$e\to \lambda e$ and introduce $\lambda$-dependent single-particle
wave functions $\Psi_{n,x}$ in these expressions. For the sake of
simplicity, we do not show explicitly the $\lambda$-dependence
generated in this manner. We find
\bea
\cC^{(2)}_{a,b}&=&
-\hf\partial_\rho\sigma(a,\rho_a) {\tilde D}_{a,b}
\partial_\rho\sigma(b,\rho_b)+\sum_{n,m=0}^\infty{1\over n!m!}
\rho^n_a\gamma^{(n+1,m+1)}_{a,b}\rho^m_b\nonu
&&+\delta_{a,b}\left(\partial^2_\rho U(a,\rho_a)
+\hf D_{a,b}\partial^2_\rho\sigma(a,\rho_a)
-\hf\partial^2_\rho\sigma(a,\rho_a)\int_c {\tilde D}_{a,c}\sigma(c,\rho_c)
+\sum_{n,m=0}^\infty{1\over n!m!}\rho^n_a
\int_c\gamma^{(n+2,m)}_{a,c}\rho^m_c\right)
\eea
in this parameterization, with ${\tilde D}_{a,b}=D_{a,b}D_{b,a}=(D_{a,b})^2$.

\subsection{Evolution equation for the density Green function}\label{matrix}
We consider first the evolution equation \eq{evolph} for arbitrary densities
 in an
$f_1c_2$ scheme for the effective action of Eq. \eq{genan}. The potential
$\cC[\rho]=\int_a U(a,\rho_a)$ is the local $f_1$ part of the approximation.
Taking the second functional derivatives of both sides of
Eq. \eq{evolph} and keeping the terms up to
the bi-local ones ($n=2$) on the r.h.s, one obtains:
\bea
\partial_\lambda (\tcGi)_{a,b}&=&2\lambda e^2 \ide_b\delta_{a,b}
-\partial_\lambda \partial_\rho^2U(a,\rho_a)\delta_{a,b}
+\lambda e^2 (\tcG\cdot \ide\cdot\tcG)_{a,a} \partial_\rho^4 U(a,\rho_a)
\delta_{a,b}\nonu
&&-\lambda e^2\left[\left(\tcG\cdot\partial_\rho^2
U\cdot\tcG\cdot\ide\cdot\tcG\right)_{a,a}
+\left(\tcG\cdot\ide\cdot\tcG\cdot\partial_\rho^2 U\cdot\tcG\right)_{a,a}
\right]\partial_\rho^4U(a,\rho_a)\delta_{a,b}\nonu
&&-\lambda e^2\left[\left(\tcG\cdot \ide\cdot\tcG\right)_{b,a}\tcG_{a,b}
+\left(\tcG\cdot \ide\cdot\tcG\right)_{a,b}\tcG_{b,a}\right]
\partial_\rho^3 U(a,\rho_a)\partial_\rho^3 U(b,\rho_b).
\eea
Separating the terms $a=b$ and $a\not= b$ on both sides and exploiting
the symmetry $\tcG_{a,b}=\tcG_{b,a}$ the following
evolution equations are obtained for the particle-hole Green function and the
local potential
\bea
\partial_\lambda\left[(\tcGi)_{a,a}-\partial_\rho^2U(a,\rho_a)\right]&=&
2\lambda e^2\left(\ide_b\delta_{a,b}\right)_{| a=b}
+\lambda e^2 (\tcG\cdot \ide\cdot\tcG)_{a,a} \partial_\rho^4 U(a,\rho_a)\nonu
&&-2\lambda e^2\left(\tcG\cdot \partial_\rho^2 U\cdot\tcG\cdot
\ide\cdot\tcG\right)_{a,a}\partial_\rho^4U(a,\rho_a)
-2\lambda e^2\left(\tcG\cdot \ide\cdot\tcG\right)_{a,a}\tcG_{a,a}
(\partial_\rho^3U(a,\rho_a))^2,\nonu
\partial_\lambda (\tcGi)_{a,b}&=&-2\lambda e^2
\left(\tcG\cdot \ide\cdot\tcG\right)_{b,a}\tcG_{a,b}
\partial_\rho^3 U(a,\rho_a)\partial_\rho^3 U(b,\rho_b).
\eea
The numerical integration of these equations with the initial condition obtained
by diagonalising the Hamiltonian for a non-interacting system represents
an algorithm whose computational requirement depends on the space-time
resolution of the Green function $\tcG$ required. The density $\rt$ can be
determined by integrating the first functional derivative of Eq. \eq{evolph}
evaluated at $\rho=\rt$ in $\lambda$,
\be
\partial_\lambda \rt_a = (\partial_\lambda \partial_\rho U \cdot \tcG)_a
   - \lambda e^2 \int_b \partial_\rho^3 U(b,\rho_b) (\tcG\cdot \ide\cdot
\tcG)_{b,b} \tcG_{b,a} .
\ee

\subsection{Evolution of the ground state ($f_2$)}
The numerically more involved but more illuminating truncation scheme is
where the quasi-particle wave functions are introduced in parametrising
the effective action. Since the quasi-particles are well defined
 for excitations
around the ground state we consider the evolution equation corresponding to
these schemes at the ground state only. Since
\be
\tr\left[\tcG\cdot\ide+\tG\cdot D\right]=-2\tr\tG\cdot\ide\cdot
\sum_{n=0}^\infty\left(\tG\cdot{\lambda^2e^2\over\Delta}\right)^{2n+1}
\ee
and
\bea
\tr G^{-1}\cdot\partial_\lambda G
&=&\beta\int_\mb{x}\Biggl[\sum_{n=1}^\infty\partial_\lambda\partial_tg_{n,n}(0)
\Psi_{n,\mb{x}}\Psi^*_{n,\mb{x}}+\sum_{n=1}^\infty\partial_tg_{n,n}(0)
\partial_\lambda\left(\Psi_{n,\mb{x}}\Psi^*_{n,\mb{x}}\right)
-\sum_{n=1}^N\partial_\lambda E_n\Psi_{n,\mb{x}}\Psi^*_{n,\mb{x}}\nonu
&&+\left(-E_n-{\hbar^2\over2m}\Delta_\mb{x}
-\mu+\ve_\mb{x}\right)\sum_{n=1}^N\partial_\lambda
\left(\Psi_{n,\mb{x}}\Psi^*_{n,\mb{x'}}\right)_{\vert\mb{x}=\mb{x'}}\Biggr],
\eea
the expressions \eq{fgrspi}, \eq{phopr} and the relation \eq{enh}
give the evolution equation as a generalized HF equation for
orthogonal wave functions,
\be\label{ghf}
{\cal R}[\Psi^*,\Psi,E]
=\sum_{n=1}^N\int_\mb{x}\left[\partial_\lambda\Psi^*_{n,\mb{x}}
{\delta{\cal H}[\Psi^*,\Psi,E,\rho]\over\delta\Psi^*_{n,\mb{x}}}
+\partial_\lambda\Psi_{n,\mb{x}}
{\delta{\cal H}[\Psi^*,\Psi,E,\rho]\over\delta\Psi_{n,\mb{x}}}
\right]_{\vert\rho=\rho_\mr{gr}}
\ee
imposed together with the auxiliary condition
\be\label{aux}
{\partial{\cal H}[\Psi^*,\Psi,E,\rho]\over
\partial E_n}_{\vert\rho=\rho_\mr{gr}} =0
\ee
where the generalized HF functional
\be\label{hffunc}
{\cal H}[\Psi^*,\Psi,E,\rho]={\cal H}^\mr{free}[\Psi^*,\Psi,E]
+{\cal H}^\mr{Coulomb}[\Psi^*,\Psi]
+{\cal H}^\mr{photon}[\Psi^*,\Psi,\rho]
+{\cal H}^\mr{int}[\Psi^*,\Psi,\rho],
\ee
is given by
\bea\label{hfham}
{\cal H}^\mr{free}[\Psi^*,\Psi,E]&=&n_s\sum_{n=1}^N\left[E_n+\int_\mb{x}
\Psi_{n,\mb{x}}^*\left(-E_n-{\hbar^2\over2m}\Delta
+\ve_\mb{x}\right)\Psi_{n,\mb{x}}\right],\nonu
{\cal H}^\mr{photon}[\Psi^*,\Psi,\rho]&=&{1\over2\beta}\tr\log
\left(-\Delta+\sigma(\rho)+\lambda^2e^2\tG\right),\nonu
{\cal H}^\mr{Coulomb}[\Psi^*,\Psi]&=&{n_s^2\lambda^2e^2\over2}
\sum_{m,n=1}^N\int_{\mb{x},\mb{y}}\Psi^*_{m,\mb{x}}\Psi_{m,\mb{x}}
{1\over4\pi|\mb{x}-\mb{y}|}\Psi^*_{n,\mb{y}}\Psi_{n,\mb{y}},\nonu
{\cal H}^\mr{int}[\Psi^*,\Psi,\rho]
&=&{1\over \beta}\rt\cdot{\delta\cC[\rho]\over\delta\rho}
-{1\over 2\beta}\left({\delta\cC[\rho]\over\delta\rho}
-\rt\cdot{\lambda^2e^2\over\Delta}\right)\cdot\tG\cdot
\left({\delta\cC[\rho]\over\delta\rho}
-{\lambda^2e^2\over\Delta}\cdot\rt\right).
\eea
and the inhomogeneous term is
\bea\label{inhhf}
\beta{\cal R}[\Psi^*,\Psi,E]
&=&2\lambda e^2\tr\tG\cdot\ide\cdot\sum_{n=0}^\infty
\left(\tG\cdot{\lambda^2e^2\over\Delta}\right)^{2n+1}
-\lambda e^2\sum_{n=1}^\infty(-1)^n\tr\left[\tcG\cdot\left(
\cC^{(2)}[\rho_\mr{gr}]\cdot\tcG\right)^n\cdot\ide\right]\nonu
&&+\lambda e^2 \tr D\cdot \tG\cdot D\cdot \sigma(\rho_\mr{gr})
\cdot \biggl[1- D\cdot\sigma(\rho_\mr{gr})\biggr]
-\hf \tr D\cdot\biggl[1- \sigma(\rho_\mr{gr})\cdot D\biggr]
\cdot \partial_\lambda \sigma(\rho_\mr{gr})\nonu
&&-\hf\left({\delta\cC[\rho_\mr{gr}]\over\delta\rho}
-\rt\cdot{\lambda^2e^2\over\Delta}\right)\cdot
\tG\cdot{2\lambda e^2\over\Delta}\cdot\tG\cdot
\left({\delta\cC[\rho_\mr{gr}]\over\delta\rho}
-{\lambda^2e^2\over\Delta}\cdot\rt\right)
-\partial_\lambda \cC_\mr{exct}[\rho_\mr{gr}]
\eea
with $\partial_\lambda\sigma[\rho_\mr{gr}]
=\partial_\lambda\sigma(\rho)_{\vert\rho=\rho_\mr{gr}}$.
Note that the auxiliary condition \eq{aux} does not determine the
value of $E_n$. Instead, the functions $g_{n,n'}(t)$ and the
constants $E_n$ appearing in the particle-hole propagator are fixed by
the dynamical relation \eq{fgrspi} for a given set of wave functions
$\Psi_n$. It is worthwhile noting that similarly to the usual
HF method the spectrum $E_n$ plays the role of  Lagrange
multipliers in the evolution equation in order to arrive at normalized
 wave functions.

Another quantity not determined by Eq. \eq{ghf} is
$\cC_\mr{exct}[\rho_\mr{gr}]$. Different choices for this
$\lambda$-dependent number lead to different wave functions $\Psi_n$
and spectrum $E_n$, different quasi-particles in short. The choice
of the basis used in the expansion \eq{sqop} of the electron field
is a central issue of any approximation scheme. The guiding
rule to define the quasi-particles is to minimize their residual
interactions, the higher order perturbative contributions. Since we
are not explicitly computing the latter we turn to another, hopefully
closely related principle for choosing the quasi-particles, the simplicity
of the evolution equation, ${\cal R}[\Psi^*,\Psi,E]=0$. The wave
functions $\Psi_n$ and the spectrum $E_n$ can then be
determined by the HF strategy, by minimizing
${\cal H}[\Psi^*,\Psi,E,\rho_\mr{gr}]$ for normalized wave functions
at each value of $\lambda$, i.e. by imposing
\be\label{hfegy}
{\delta{\cal H}[\Psi^*,\Psi,E,\rho]\over\delta\Psi^*_{n,x}
}_{\vert\rho=\rho_\mr{gr}}
={\delta{\cal H}[\Psi^*,\Psi,E,\rho]\over\delta\Psi_{n,x}
}_{\vert\rho=\rho_\mr{gr}}=0.
\ee
One can construct the renormalization group flow with the
accuracy of $\ord{\delta\lambda}$ when the wave functions at
$\lambda+\delta\lambda$ are found by minimizing the functional
${\cal H}[\Psi^*,\Psi,E,\rho_\mr{gr}]$ computed at $\lambda$.
The initial value for $\rho_\mr{gr}$ is $\rho_\mr{gr} (\lambda =0) =
{\rt}^{(0)}$, the ground-state density of the non-interacting system.

The simplest truncation is to retain the terms $\ord{\rho^2}$ and $\ord{e^2}$
of the density functional which gives $\cC[\rho]=$const. and
$\sigma(\rho)=0$
according to the leading order expression \eq{densfm}, a $c_2$ scheme.
The dynamical problem, the evolution of the matrix elements
\eq{fgrspi} does not appear at this level and the remaining
static (time independent) evolution equation
realizes the usual HF scheme following what happens
when the interaction strength $e$ in $H_2$ is turned on gradually.
The evolution offers in this manner the possibility of approaching the
true HF minimum in infinitesimal, perturbative steps.
Our original assumption about the smoothness of the $e$-dependence
requires in the one-loop approximation the absence of local minima
of the HF energy along the path $0<\lambda<1$
which would prevent us to find the true minimum.
The terms $\ord{\rho^3}$ and $\ord{e^4}$ ignored in the HF
equation represent the multi-body forces and the contributions
of the vacant HF orbits. But even the truncation $\cC=$const.
and $\sigma(\rho)=0$ gives an already improved HF scheme
where the exchange interaction coming
from the second line of Eq. \eq{hfham} and the energy of the
zero point fluctuations when ${\cal R}=0$ is solved contain
a partial resummation of the perturbation expansion in $e$. The dynamical
HF functional given by Eq. \eq{hffunc} implies the time-dependence of
the propagator due to the intermediate states appearing in the
higher orders of the perturbation expansion.

We find it remarkable that the
main feature of the HF scheme, the determination of the structure
of the quasi-particles making up the ground state by minimizing
a functional, can be maintained when higher order radiative corrections
are retained. But there are complications which can easiest
be understood by adopting the language of the perturbation
expansion and recalling that the higher order contributions
contain excitations around the unperturbed ground state.
These contributions correspond to excitations and de-excitations
of the ground state and imply time-dependence. As a result, despite that
the original HF functional used in the determination of the
ground state is static and refers to the spatial dependence of the
wave functions only, the one given by Eq. \eq{hffunc} involves dynamical,
time-dependent quantities.
Furthermore, the construction of the intermediate excited states
appearing in the higher orders of the perturbation expansion
is achieved by the simultaneous solution of the evolution
equation for the `potential energy' of the density functional
which is mainly responsible for the excitations.

\subsection{Evolution of the potential energy ($f_2$)}
The generalized HF equation does not determine the
evolution of the potentials $U(\rho)$ and $\sigma(\rho)$ and that of
the matrix $\gamma^{(n,m)}$. In order
to find them, we have to consider the effective action in its
full bi-local form,
\bea
&&\hf(\rho-\rt)\cdot\partial_\lambda\tcGi\cdot(\rho-\rt)
-\partial_\lambda\rt\cdot\tcGi\cdot(\rho-\rt)
+\int_a\partial_\lambda U(a,\rho_a)\nonu
&&+\sum_{n,m=0}^\infty{1\over2n!m!}\rho^n\cdot
\partial_\lambda\gamma^{(n,m)}\cdot\rho^m
+\partial_\lambda\left(-n_s\tr\log G^{-1}+\hf\tr\log[-D^{-1}\Delta]
+\hf\tr\ D\cdot\sigma-{1\over4}\tr\ \sigma\cdot D\cdot\sigma\cdot D\right)\nonu
&=&-\lambda e^2\left(\tr\ide\cdot\tcG-\tr\ide\cdot\tcG\cdot
\cC^{(2)}\cdot\tcG+\tr\ide\cdot\tcG\cdot
\cC^{(2)}\cdot\tcG\cdot\cC^{(2)}\cdot\tcG+\cdots\right).
\eea
We introduce the parameterization $\rho=\rho_\mr{gr}+\eta$ and
use the identity
\bea\label{evdiff}
\partial_\lambda\left[\Gamma[\rho] 
+{\lambda^2e^2\over2}\rho\cdot\ide\cdot
\rho\right]_{\vert\rho=\rho_\mr{gr}+\eta} 
&=&\partial_\lambda\left[\Gamma[\rho]
+{\lambda^2e^2\over2}\rho\cdot\ide\cdot
\rho\right]_{\vert\rho=\rho_\mr{gr}}
+\partial_\lambda(\cC[\rho_\mr{gr}+\eta]-\cC[\rho_\mr{gr}])\nonu
&&+\eta\cdot\partial_\lambda\tcGi\cdot(\rho_\mr{gr}-\rt)
+\hf\eta\cdot\partial_\lambda\tcGi\cdot\eta
-\partial_\lambda\rt\cdot\tcGi\cdot\eta
\eea
to write the difference of the evolution equations
imposed at $\eta\not=0$ and $\eta =0$ as
\bea\label{ubilev}
&&\lambda e^2\sum_{n=1}^\infty(-1)^n\tr\left\{\tcG\cdot\left[\left(
\cC^{(2)}[\rho_\mr{gr}]\cdot\tcG\right)^n
-\left(\cC^{(2)}[\rho_\mr{gr}+\eta]\cdot\tcG\right)^n
\right]\cdot\ide\right\}\nonu
&&=\partial_\lambda(\cC[\rho_\mr{gr}+\eta]-\cC[\rho_\mr{gr}])
+\hf\eta\cdot\partial_\lambda\tcGi\cdot\eta
+\eta\cdot\partial_\lambda\tcGi\cdot(\rho_\mr{gr}-\rt)
-\partial_\lambda\rt\cdot\tcGi\cdot\eta.
\eea
Note that there is no term with $\partial_\lambda\rho_\mr{gr}$
in this expression because the substitution $\rho\to\rho_\mr{gr}+\eta$
is made after the evaluation of the derivative $\partial_\lambda$
in Eq. \eq{evdiff}. As mentioned earlier, the inequality
$|\eta|<\rho_\mr{gr}$ is not required.

Eq. \eq{ubilev} together with the evolution of the ground state
will be studied below in a more detailed manner in the ultra-local
and the bi-local levels for the density fluctuation $\eta$.
We shall consider the potentials first and the results obtained
will be used to find the explicit evolution of the ground state.

\subsection{Ultra-local approximation ($f_1c_2$)}\label{ultrls}
The scheme $c_2$ is based on the ansatz \eq{genan} with
$\cC[\rho]=\Gamma_0$ being a $\lambda e$ dependent but $\rho$-independent quantity.
The effective actions of such form make up the
space ${\cal A}_{c_2}$. This scheme will be improved by introducing
an ultra-local part containing the functions
$U(x,\rho_x)$ and $\sigma(x,\rho_x)$ which have no linear piece
in the Taylor expansion around $\rho=0$. This is achieved by
keeping the ultra-local pieces without linear term
only in Eqs. \eq{cqfl} and \eq{vans}. The ultra-local component
of Eq. \eq{ubilev} restricted into the orthogonal
subspace ${\cal A}_{f_1}\cap{\cal A}_{c_2}^\bot$ can be written as
\be\label{diffi}\label{ulev1f2c}
\partial_\lambda\psi_{\mr{gr}a}(\lambda,\eta)
=f_a(\lambda)\partial_\eta^2\psi_{\mr{gr}a}(\lambda,\eta),
\ee
for the function
\be\label{lincomb}
\psi_a(\lambda,\rho)={D_{a,a}\over2}\sigma(a,\rho)+U(a,\rho),
\ee
where $D_{a,a}$ is rendered finite by the UV cut-off,
\be
f_a(\lambda)=\lambda e^2\left(\tcG\cdot\ide\cdot\tcG\right)_{a,a}, 
\ee  
and the notation $f_\mr{gr}(\eta)=f(\rho_\mr{gr}+\eta)-f(\rho_\mr{gr})$
was introduced to denote functions of the density measured from the
ground state value.

Two remarks are in order concerning Eq. \eq{diffi}. The parameter
$\lambda$ controls the amplitude of quantum fluctuations  
in the internal space and therefore it is rather natural that the
evolution equation is a diffusion equation in the internal space  
with $\lambda$ as time. Furthermore, this equation is linear for
the combination \eq{lincomb} and does not determine the potential
$U(\rho)$ and the function $\sigma(\rho)$ independently.

By assuming the initial condition $\sigma=U=\psi=0$ we have no
evolution because Eq. \eq{ulev1f2c} is linear and homogeneous.
The constrained bi-local form of the effective action is sufficiently
general and further renormalization takes place only due to the
three-body interactions, at the tri-local level.

The evolution of the ground state energy is rather simple since
$\cC^{(2)}=\partial_\rho^2\psi=0$. In fact, $\Gamma_0$ is
found by integrating out the differential equation
\be\label{cevhf}
\partial_\lambda\Gamma_0
=2\lambda e^2\tr\tG\cdot\ide\cdot\sum_{n=0}^\infty
\left(\tG\cdot{\lambda^2e^2\over\Delta}\right)^{2n+1}
-\hf\rt\cdot{\lambda^2e^2\over\Delta}\cdot
\tG\cdot{2\lambda e^2\over\Delta}\cdot\tG\cdot
{\lambda^2e^2\over\Delta}\cdot\rt,
\ee
where the quasiparticle wave functions $\Psi$ on the right hand side
are obtained by solving the generalized HF equation Eq. \eq{hfegy}.
The leading order contribution of the photon determinant gives
the HF exchange term and its remaining contributions together
with the terms of $\ord{e^4}$ in $\Gamma_0$
represent the corrections to the traditional HF result.

It is instructive to follow what happens if the evolution equation is
allowed to adjust the linear part of the local potential what has been
fixed in the computation above by the first term of the ansatz
\eq{genan}. One can see that the ultra-local part
of the effective action is then modified in such a manner that
the density in the ground state as given by Eq. \eq{onegr}
involves the partially resummed particle-hole propagator.
The improvement is that we have a partial resummation of the
leading order self-energy insertion. This is in agreement with
the observation that the integration of the leading order
renormalization group equation resums the leading order
self-energy insertions in the propagators \cite{resum}, \cite{coll}.
The weakness of this solution compared with the HF level is
that the non-interacting quasi-particles are used in the construction
of the 'free' particle-hole propagator. This happens because
we did not let the bi-local part of the evolution equation
to correct for the correlations in the ground state and left the
ultra-local part alone to determine the ultra-local potential
and with it the location of the minimum in the space of
density configurations $\rho_x$.

\subsection{Bi-local approximation ($f_2$)}
The bi-local part of the evolution equation \eq{ubilev} reads as
\bea
&&\sum_{n,m=0}^\infty{\partial_\lambda\gamma^{(n,m)}_{a,b}\over2n!m!}
[(\rho_\mr{gr}+\eta)^n_a(\rho_\mr{gr}+\eta)^m_b
-\rho_{\mr{gr}a}^n\rho_{\mr{gr}b}^m]
-{1\over4}\partial_\lambda\left[\tilde D_{a,b}\left( 
\sigma(\rho_a+\eta_a)\sigma(\rho_{b}+\eta_b) 
-\sigma(\rho_{a})\sigma(\rho_{b})\right)\right]_{\rho=\rho_\mr{gr}}\nonu
&&=\lambda e^2\int_{c,d}\ide_{c,d}\tcG_{d,a}\tcG_{b,c}\Biggl\{ 
\sum_{n,m=0}^\infty{\gamma^{(n+1,m+1)}_{a,b}\over n!m!}
[(\rho_\mr{gr}+\eta)^n_a(\rho_\mr{gr}+\eta)^m_b
-\rho_{\mr{gr}a}^n\rho_{\mr{gr}b}^m]\nonu
&&-\hf\tilde D_{a,b}[\partial_\eta\sigma(a,\rho_{\mr{gr}a}+\eta_a) 
\partial_\eta\sigma(b,\rho_{\mr{gr}b}+\eta_b)
-\partial_\rho\sigma(a,\rho_{\mr{gr}a})
\partial_\rho\sigma(b,\rho_{\mr{gr}b})]\nonu
&&-[\partial^2_\eta\psi_a(\lambda,\rho_{\mr{gr}a}+\eta_a)
\partial^2_\eta\psi_b(\lambda,\rho_{\mr{gr}b}+\eta_b)
-\partial^2_\rho\psi_a(\lambda,\rho_{\mr{gr}a})
\partial^2_\rho\psi_b(\lambda,\rho_{\mr{gr}b})]\tcG_{a,b}\nonu
&&-\hf\delta_{a,b}\int_e\tilde D_{a,e}
[\partial_\eta^2\sigma(a,\rho_{\mr{gr}a}
+\eta_a)\sigma(e,\rho_{\mr{gr}e}+\eta_e)
-\partial_\rho^2\sigma(a,\rho_{\mr{gr}a})\sigma(e,\rho_{\mr{gr}e})]\nonu
&&+\delta_{a,b}\sum_{n,m=0}^\infty\int_e{\gamma^{(n+2,m)}_{a,e}\over n!m!}
[(\rho_\mr{gr}+\eta)^n_a(\rho_\mr{gr}+\eta)^m_e 
-\rho_{\mr{gr}a}^n\rho_{\mr{gr}e}^m]\Biggr\}
- \hf  \partial_\lambda \tcGi_{a,b}\eta_a \eta_b . 
\eea
After taking into account the ultra-local solution $\sigma=\psi=0$
and projecting out the bi-linear $\rho$-dependence present
already in the $\ord{\rho^2}$ $c_2$-part of the effective action
this equation reduces to
\bea
&&\sum{\partial_\lambda\gamma^{(n,m)}_{a,b}\over2n!m!}
[(\rho_\mr{gr}+\eta)^n_a(\rho_\mr{gr}+\eta)^m_b
-\rho_{\mr{gr}a}^n\rho_{\mr{gr}b}^m]\nonu
&&=\lambda e^2\int_{c,d}\ide_{c,d}\tcG_{d,a}\tcG_{b,c}\Biggl\{
\sum{\gamma^{(n+1,m+1)}_{a,b}\over n!m!}
[(\rho_\mr{gr}+\eta)^n_a(\rho_\mr{gr}+\eta)^m_b
-\rho_{\mr{gr}a}^n\rho_{\mr{gr}b}^m]\nonu
&&+\delta_{a,b}\sum\int_e{\gamma^{(n+2,m)}_{a,e}\over n!m!}
[(\rho_\mr{gr}+\eta)^n_a(\rho_\mr{gr}+\eta)^m_e
-\rho_{\mr{gr}a}^n\rho_{\mr{gr}e}^m]\Biggr\}
\eea
where the summation over $n,m=2,\cdots,\infty$, $(n,m)=(1,2)$ and $(2,1)$.
Since this is a linear, homogeneous differential equation and the
initial condition is $\gamma^{(n,m)}=0$, the solution is identically
vanishing. Together with the similarly negative result
of the ultra-local evolution one concludes that corrections
to the scheme $c_2$ are generated at three-body forces only.

\subsection{Exchange energy in the $f_1c_2$ approximation}
It was noted above that the exchange energy comes from
the photon fluctuation determinant. The evolution
in the strength of the Coulomb interaction 'turns' the photon
dynamics gradually on and should by itself generate the 
photon contributions to the effective action. Thereby, one
wonders if the evolution can generate most of the exchange interaction
without our choice of the ansatz for the density functional, i.e.
by neglecting the particle-hole contribution to the photon
propagator, $D\to-1/\Delta$. Let us simplify matters by
restricting ourselves to a constrained bi-local approximation,
$\gamma^{(m,n)}=0$.

We note first that the basic idea of the KS scheme, the
trade off the exchange energy for potential energy is nicely
realized by the approximation $f_1$ of Section \ref{ultrls}.
In fact, the evolution equation in this
approximation is invariant with respect to the transformation
\be\label{ksinv}
U(a,\rho)\to U(a,\rho)+\epsilon(a,\rho),\ \
\sigma(a,\rho)\to\sigma(a,\rho)-{2\over D_{a,a}}\epsilon(a,\rho).
\ee
In the absence of the density dependence in the exchange
interaction, as noted in section \ref{exchi} the
success of the KS scheme can be ascribed to the efficiency of the
ansatz where a local potential is kept, $U(\rho)\not=0$ in the
absence of exchange interactions, $D^{-1}=-\Delta$,
\be\label{ksefa}
\Gamma[\rho]=\hf(\rho-\rt)\cdot\tcGi\cdot(\rho-\rt)
-{\lambda^2e^2\over2}\rho\cdot\ide\cdot\rho
-n_s\tr\log G^{-1}+\Gamma_0.
\ee
The evolution equation,
\bea
&&-\partial_\lambda\rt\cdot\tcGi\cdot(\rho-\rt)
+\partial_\lambda\Gamma_0-n_s\partial_\lambda\tr\log G^{-1}\nonu
&=&-\hf(\rho-\rt)\cdot\partial_\lambda\tcGi\cdot(\rho-\rt)
-\lambda e^2\sum_{n=0}^1(-1)^n\tr\left[\tcG\cdot\left(
\cC^{(2)}[\rho]\cdot\tcG\right)^n\cdot\ide\right].
\eea
considered at the minimum of the effective action, $\rho=\rho_\mr{gr}$
simplifies to
\bea
-\partial_\lambda\rt\cdot{\lambda^2e^2\over\Delta}\cdot\rt
-n_s\partial_\lambda\tr\log G^{-1}+\partial_\lambda\Gamma_0
&=&-\lambda e^2\tr\ \tcG\cdot\ide
+\partial_\lambda\rt\cdot{\lambda^2e^2\over\Delta}\cdot\tG
\cdot{\lambda^2e^2\over\Delta}\cdot\rt\nonu
&&-\hf\rt\cdot{\lambda^2e^2\over\Delta}\cdot
\left(\tG\cdot{2\lambda e^2\over\Delta}\cdot\tG
-\partial_\lambda\tG\right)\cdot{\lambda^2e^2\over\Delta}\cdot\rt.
\eea
because $\cC^{(2)}=0$. We can write this equation as
\be\label{gsk}
\sum_{n=1}^N\int_\mb{x}\left[\partial_\lambda\Psi^*_{n,\mb{x}}
{\delta{\cal H}_\mr{KS}[\Psi^*,\Psi,E,\tilde\rho]
\over\delta\Psi^*_{n,\mb{x}}}
+\partial_\lambda\Psi_{n,\mb{x}}
{\delta{\cal H}_\mr{KS}[\Psi^*,\Psi,E,\tilde\rho]
\over\delta\Psi_{n,\mb{x}}}
\right]_{\vert\tilde\rho=\rt}=0
\ee
together with the auxiliary condition \eq{aux}. The generalized 
HF functional
\be
{\cal H}_\mr{KS}[\Psi^*,\Psi,E,\tilde\rho]=
{\cal H}^\mr{free}[\Psi^*,\Psi,E]
+{\cal H}^\mr{Coulomb}_\mr{KS}[\Psi^*,\Psi,\tilde\rho]
+{\cal H}^\mr{int}_\mr{KS}[\Psi^*,\Psi],
\ee
is quadratic in the wave functions in the leading order of $e$,
\bea\label{KSfunc}
{\cal H}_\mr{KS}^\mr{Coulomb}[\Psi^*,\Psi,\tilde\rho]&=&{n_s\lambda^2e^2}
\int_{\mb{x},\mb{y}}\sum_{n=1}^N\Psi^*_{n,\mb{x}}\Psi_{n,\mb{x}}
{1\over4\pi|\mb{x}-\mb{y}|}\tilde\rho_{\mb{y},0}\\
{\cal H}^\mr{int}_\mr{KS}[\Psi^*,\Psi,\rho]
&=&-{1\over 2\beta}\rt \cdot{\lambda^2e^2\over\Delta}\cdot\tG\cdot
{\lambda^2e^2\over\Delta}\cdot\rt.
\eea
and $\Gamma_0$ is given by the differential equation
\bea\label{cev}
\partial_\lambda\Gamma_0&=&-\lambda e^2 \tr\ \tG\ide
-\hf\rt\cdot{\lambda^2e^2\over\Delta}
\cdot\tG\cdot{2\lambda e^2\over\Delta}\cdot\tG\cdot
{\lambda^2e^2\over\Delta}\cdot\rt
\eea
together with the initial condition $\Gamma_0(\lambda=0)=0$.

The solution of the evolution equation is given by the wave functions
at the minimum of the functional ${\cal H}_\mr{KS}$, the solution of the
HF equations \eq{hfegy} together with the auxiliary condition
\eq{aux} obtained by the replacement ${\cal H}_\mr{HF}\to{\cal H}_\mr{KS}$.
The first term on the right hand side of Eq. \eq{cev} is canceled
in Eq. \eq{cevhf} against the leading order contribution of the
photon determinant, the HF exchange term. In the present ansatz
there is no exchange term but the evolution equation generates one since
\be
\Gamma_0=-{\lambda^2e^2\over2}\tr\ \tG\ide+\ord{e^4}.
\ee

In DFT one separates the kinetic energy $T[\rho]$ and the direct
Coulomb contribution by writing
\be\label{ksf}
E_v[\rho]=T[\rho]+\hf\int_{\mb{x},\mb{y}}
\rho_\mb{x}{e^2\over4\pi|\mb{x}-\mb{y}|}\rho_\mb{y}+E_{xc}[\rho],
\ee
and letting $E_{xc}[\rho]$ for the exchange contributions and
radiative corrections. The effective action $\Gamma^\mb{free}[\rho]$
introduced above for the non-interacting fermion problem gives
according to Eqs. \eq{gafree} and \eq{gaexfr}
\be
T[\rho]=\lim_{\beta\to0}{1\over\beta}\Gamma^\mb{free}[\rho]
=\lim_{\beta\to0}{1\over\beta}\left[\Gamma^\mb{free}_0
+\hf(\rho-\rt)\cdot \tGi \cdot(\rho-\rt)\right]+\ord{(\rho-\rt)^3}.
\ee

The exchange and correlation functional $\Gamma_{xc}[\rho]$, the
dynamical generalization of $E_{xc}[\rho]$, introduced in
Eq. \eq{ksf} for static density only, is defined as
\be\label{gaxc}
\Gamma_\mr{xc}[\rho]=\Gamma [\rho]-\Gamma^\mb{free}[\rho]
+{e^2\over2}\rho\cdot\ide\cdot\rho.
\ee
In the approximation $f_1c_2$ we find
\be\label{gaxce}
\Gamma_\mr{xc}[\rho]=\hf(\rho-\rt)\cdot\left(\tGi+{e^2\over\Delta}\right)
\cdot(\rho-\rt)-n_s\tr\log G^{-1}G^{(0)}+\Gamma_0,
\ee
c.f. Eq. \eq{ksefa}, where $G^{(0)}$ is the fermion propagator in the
non-interacting system.

It is important to note that although the effective action
formalism is in contradiction with the basic assumption of the
KS scheme this latter can be realized in a given truncation
of the density functional. In fact, it has been noted
that the exchange correlation, the one-loop contribution
to the photon path integral, represented by $\Gamma_0$ in Eq.
\eq{gaxce} is independent of the density. But an artificially introduced
density-dependence of the photon self-energy can be traded against
a local density-dependent potential in the approximation $f_1$
and opens the possibility of realizing the KS strategy.

\section{Summary}
A renormalization group motivated method is introduced to perform
the resummation of the perturbation expansion to the effective
action for the density, the density functional, by solving
a functional differential equation for the effective action.
It is shown that this evolution equation
which corresponds to the gradual increase of the
strength of the Coulomb interaction generates the electrostatic
energy of the charged particle system. In practical applications
one has to project this equation on a restricted subspace of
the effective action functionals and a systematic approximation
scheme is suggested by truncating the functionals according to their
multi-local structure.

The simplest non-trivial truncation level reproduces the
traditional HF scheme. The radiative corrections on the bi-local
level where two-body correlations are retained only are captured
by our generalized HF equations. The more
involved approximation schemes provide a systematic generalization
of the HF method by keeping the three-body and higher correlations.
The most important complexity of these improved
schemes comes from the fact that the higher order corrections
of the perturbation expansion include contributions from virtual
excitations which in turn renders the functionals in consideration
dynamical, i.e. containing the time-dependence of the density as
the independent variable.
The generalized HF equation determines the wave functions of the
 quasi-particles making up the ground
state by simultaneously minimizing the dynamical HF functional
and solving the evolution equation for other functions
characterizing the effective action.

Generator functionals offer a formal way to separate
the direct and the exchange energy contributions according to the
level they enter in the multi-local expansion of the functionals
or the saddle point expansion of the photon path integral.
The direct and the exchange terms are responsible for
the local, classical and the bi-local, one-loop contributions,
respectively.

The exchange term identified in this manner is density independent,
in contrast to the basic assumption of the KS scheme. But it is
shown that the claimed density dependence of the correlation energy
can be traded into a local potential, independent of the
structure of the quasi-particles in a given approximation
of the density functional. The evolution does in fact generate in this
case the correct exchange contribution to the ground state energy
and the corresponding generalized HF equation gives an improved KS scheme.
An explicit expression is obtained for the exchange-correlation
functional.

Finally, we mention two possibilities concerning the
choice of the quantum state considered.
According to the Wick's theorem, the expectation value
$\la\Psi|A|\Psi\ra$ of the operator $A$ depends on the state
$|\Psi\ra$ by the choice of the non-interacting path integral, the free
propagator, and the Feynman rules which give account the interactions
in a manner independent of $|\Psi\ra$. In our scheme
the interactions determine the right hand side of the evolution
equation \eq{seveq} and the state $|\Psi\ra$ is reflected in the
parametrization of the electron propagator. One possibility is
to generalize the ground state which according
to the ansatz \eq{fgrspi} corresponds to a single Slater determinant.
By the introduction of an over-complete, non-orthogonal basis
$\Psi_n$ and the parameterization $G_{x,x'}=\sum_{n,n'=1}^\infty 
g_{n,n'}(t-t')\Psi_{n,\mb{x}}\Psi^*_{n',\mb{x'}}$
we can mix several Slater determinants in the ground state.
Another remark is about the issue of excited states.
Suppose that the generalized HF equations \eq{hfegy} have been solved,
a set of wave functions $\Psi_n$ has been obtained and $N$ of them,
with the lowest `eigenvalues' $E_n$ have been used to construct
the ground state. The repetition of these steps when
some of these wave functions are replaced with others which
correspond to higher $E_n$ in the construction of the electron
propagator \eq{fgrspi} leads to the systematic construction of
excited states.

We believe that the general setting of this computational
algorithm for the density functional presented here,
the gradual turning up the Coulomb interaction and a
systematical approximation scheme by considering richer
functional classes in solving the evolution equation
is useful for strongly correlated electronic systems
in Solid State or Condensed Matter Physics, as well as in
Quantum Chemistry. The numerical implementation of the scheme
of Section \ref{matrix} to test our ideas is in progress.

\acknowledgments
We thank Abdel Kenoufi for a useful discussion and Dmitrij V. Shirkov
for remarks about early references on the renormalization group method.
This work has been supported by the grants OTKA T29927/98, OTKA T032501/00,
NATO SA(PST.CLG 975722)5066, and M\"OB-DAAD 27 (323-PPP-Ungarn).
One of the authors (K.S.) has been supported by the Alexander von Humboldt-Foundation
and thanks W. Greiner for his kind hospitality.

\appendix
\section{Effective action}\label{effapex}
In this Appendix we briefly review the interpretation of the 
effective action for a generic scalar field theory.

Suppose that the basic field variable of the system considered is
$\phi_x$ and the connected Green functions are given by the 
generator functional $W[j]$ expressed in terms of the path integral
\be 
e^{W[j]}=\int\cd{\phi}e^{-S_E[\phi]+\int_x\phi_xj_x}
\ee
where $S_E[\phi]$ denotes the imaginary time, Euclidean action. The
effective action $\Gamma[\Phi]$ is defined by the Legendre transformation
\be
\Gamma[\Phi]=-W[j]+\int_x\Phi_xj_x,\ \ \
\Phi_x={\delta W[j]\over\delta j_x}.
\ee
The inverse variable transformation,
\be
j_x={\delta\Gamma[\Phi]\over\delta\Phi_x}.
\ee
shows that the minimum of the effective action gives the
free energy in the absence of the source.

In order to see what happens with non-vanishing source, we consider
\be
e^{-\Gamma[\Phi]}=\max_{j_x}
\int\cd{\phi}e^{-S_E[\phi]+\int_x(\phi_x-\Phi_x)j_x}
\ee
where the source $j_x$ is chosen by maximizing the right hand side.
The field $\phi_x$ fluctuates in
the path integral around $\phi^\mr{cl}_x$, the maximum of the
integrand which satisfies the classical equation of motion,
\be\label{eqmoj}
{\delta S_E[\phi^\mr{cl}]\over\delta\phi_x}=j_x.
\ee
By ignoring the fluctuations we have
\be
\Gamma[\Phi]=\min_{j_x,\phi}
\left(S_E[\phi]+\int_x(\phi_x-\Phi_x)j_x\right)=\min_{j_x}
\left(S_E[\phi^\mr{cl}]+\int_x(\phi^\mr{cl}_x-\Phi_x)j_x\right),
\ee
where we should bear in mind that $\phi^\mr{cl}$ depends on the source
$j$. The minimization with respect $j$ yields
$\Phi=\phi^\mr{cl}$ and $\Gamma[\Phi]=S_E[\Phi]$ showing
that on the tree-level where the quantum/thermal fluctuations
are ignored the effective action reproduces the classical one.

In order to assess the role of fluctuations, let us assume that
the system is placed into a large but finite spatial volume, introduce a
basis $\phi_{n,x}$, $\int_x\phi_{n,x}\phi_{m,x}=\delta_{n,m}$,
$n=1,2,\ldots$ for a given $\Phi_x$ where the
element $n=1$ agrees with $\Phi_x$, $\Phi_x=C\phi_{1,x}$ and write
the path integral as integration over the coefficient $c_n$ in
the expansion $\phi_x=\sum_nc_n\phi_{n,x}$,
\be
e^{-\Gamma[\Phi]}=\max_{j_k}\prod_{n=1}^\infty\int dc_ne^{-S_E[c]
+(c_1-C)j_1+\sum_{n=2}^\infty c_nj_n},
\ee
We define the norm for the field configurations as
$||\phi||^2=\int_x\phi^2_x$. Clearly $|C|=||\Phi||$, indicating
that the typical values in the path integral are $c_n=\ord{||\Phi||^0}$
for $n=2,3,\ldots$ and $c_1$ takes values around
$\bar c_1=\ord{||\Phi||}$ with fluctuations $\delta c_1=\ord{||\Phi||^0}$.
We see in this manner that the relative fluctuations of the modes
$c_n$ in the path integral are weakly influenced by the source
in directions orthogonal to the desired field expectation value
$\Phi_x$ and are suppressed along the direction of $\Phi_x$ as
$\ord{||\Phi||^{-1}}$. As $j$ is chosen such that $\Phi_x$ extends 
over a larger volume the effective action tends to the free energy 
computed by integrating over configurations which have a
fixed projection on $\Phi$.

\section{Free fermions}\label{free}
We re-derive in this Appendix the free energy of non-interacting fermions
in the framework of Grassmannian path integration with
special attention payed to the continuum limit, $\Delta t\to 0$.

\subsection{Grassmannian path integral}
The fermionic amplitudes are usually obtained by integrating
over Grassmann variables \cite{grassm} $\psi_1\psi_2+\psi_2\psi_1=0$.
The value of the integral $\int\cD{\psi}e^{-S[\psi]}$ is defined in such a manner
that the equation of motion is satisfied on the average,
$\int\cD{\psi}e^{-S[\psi]}\delta S[\psi]/\delta\psi=0$. This amounts to the rule
\be\label{grint}
\int d\psi f(\psi)={df(0)\over d\psi}
\ee
since any function $f(\psi)$ of a Grassmann variable
$\psi$ can be written as $f(\psi)=\eta_1\psi+\eta_2$ where $\eta_1$
 and $\eta_2$
are Grassmann variables. One introduces pairs of complex conjugate variables
$\psid_k,\psi_k$ and the integral
\be\label{grga}
I=\prod_k\int d\psi_kd\psid_ke^{\psid_kA_{k,\ell}\psi_\ell}
={\mathrm{Det}}A
\ee
can be easily checked by expanding the exponential function of
the integrand and applying the rule \eq{grint}.

Consider now a two-level system whose wave function can be written as
$\Psi(\psid)=a+b\psid=|a,b\ra$, where $a$ and $b$ are complex numbers.
The complex conjugate will be given by
$\Psi^*(\psi)=a^*+b^*\psi=\la a,b|$. The states $|1,0\ra$ and $|0,1\ra$
are orthogonal and of unit norm by using the scalar product
\be
\la\Psi|\Phi\ra=\int d\psid d\psi e^{-\psid\psi}\Psi^*(\psi)\Phi(\psid).
\ee
We are looking for the matrix elements of the time evolution operator
\be
\la a_f,b_f|U(t)|a_i,b_i\ra
=\la a_f,b_f|e^{-i{t\over\hbar}E\psid\psi}|a_i,b_i\ra.
\ee
For this end we divide the time interval $[0,t]$ into $N$ parts,
$t= N\Delta t$, introduce a
pair of variables $\psid_k,\psi_k$ at each division point, and inspect
the integral
\bea\label{repi}
I_R&=&\prod_{k=0}^N\int d\psid_kd\psi_k\Psi_f^*(\psi_{N+1})\Psi_i(\psid_0)
e^{\sum_{j=0}^N[-\psid_j(\psi_j-\psi_{j-1})
-{i\over\hbar}E\dt\psid_j\psi_{j-1}]}\nonu
&=&\prod_{k=0}^N\int d\psid_kd\psi_k\Psi_f^*(\psi_{N+1})\Psi_i(\psid_0)
\prod_{j=0}^Ne^{-\psid_j\psi_j}\cdot
e^{(1-{i\over\hbar}E\dt)\psid_j\psi_{j-1}}.
\eea
by expanding the exponential functions. The non-vanishing contributions
proportional with $a_i$ must pick up $\psi_0$ from the
first exponential function. This contribution is independent
of $E$ and will be responsible for the propagation of the ground state.
In fact, always a pair $\psi_j\psid_j$ what comes from the expansion
of this exponential and $\la0,1|U(t)|1,0\ra=0$.
The contributions containing $b_i$ must find $\psi_0$
from the second exponential, responsible for the
propagation of the excited state. We find in this manner
\be
U(t)=\lim_{N\to\infty}\pmatrix{\left(1-{i\over\hbar}E\dt\right)^N&0\cr0&1}
=\pmatrix{e^{-{i\over\hbar}Et}&0\cr0&1}
\ee
in the basis
\be
|0\ra=|1,0\ra=\pmatrix{0\cr1},~~~~|1\ra=|0,1\ra=\pmatrix{1\cr0}.
\ee
Notice that in order to have the right sign for
the matrix element $\la1,0|U(t)|1,0\ra$ we have to impose
$\psi_{-1}=-\psi_N$, $\psid_{-1}=-\psid_N$. The matrix elements of the
imaginary time evolution
operator  are obtained in a similar manner,
\be\label{impi}
I_I=\prod_{k=0}^N\int d\psid_kd\psi_k\Psi_f^*(\psi_{N+1})\Psi_i(\psid_0)
e^{\sum_{j=0}^N[-\psid_j(\psi_j-\psi_{j-1})-E\dt\psid_j\psi_{j-1}]}.
\ee

\subsection{Continuum limit}
The naive continuum limit, $\dt\to0$ suggests
\be\label{nret}
\la\Psi_f|e^{-i{t\over\hbar}E\psid\psi}|\Psi_i\ra=
\int\cd{\psi}\cd{\psid}e^{{i\over\hbar}\int_t
\psid_t[i\hbar\partial_t-E]\psi_t}
\ee
and
\be\label{nimt}
\la\Psi_f|e^{-\beta E\psid\psi}|\Psi_i\ra=
\int\cd{\psi}\cd{\psid}e^{-\int_t\psid_t[\partial_t+E]\psi_t}.
\ee
The resulting continuum expressions are simpler than the
discrete equations but this simplification is rather misleading.
In fact, let us evaluate the continuum expression \eq{nimt} by means
of \eq{grga}. The eigenmodes of the quadratic form in the exponent
are
\be
\psi_t=e^{-i\omega t}\psi_\omega
\ee
and the corresponding eigenvalue is
\be\label{spectr}
\lambda^C_\omega=i\omega-E,
\ee
where $\omega=(2n+1)\pi/\beta$ due to the anti-periodic boundary condition
in time. We have in this manner
\bea
\mathrm{Det}[-\partial_t-E]&=&\prod_{n=-\infty}^\infty
(E-(2n+1)i\pi T)
=\prod_{k=\infty}^\infty((2k+1)i\pi T)
\prod_{n=-\infty}^\infty\left[1+{E\over(2n+1)i\pi T}\right].
\eea
By means of the relation \cite{grry}
\be
\prod_{\ell=-\infty}^\infty\left[1+{x\over(2\ell+1)i\pi}\right]
=\cosh{x\over2},
\ee
we find
\bea
{\mathrm Det}[-\partial_t-E]&=&C(T)\cosh{E\beta\over2}
=\hf C(T)e^{E\beta\over2}\left(1+e^{-E\beta}\right).
\eea
One might interpret the factor $exp(E\beta/2)$ as the result of
some kind of the zero point fluctuations and interpret the rest
as the partition function for the two level system. But this
is unacceptable because the first equation shows that the suppression
is actually symmetric in $E\to-E$. We ended up with such an
unacceptable result because the spectrum \eq{spectr} is
for continuous time, when $\dt=0$. The expressions \eq{repi},
\eq{impi} were obtained for $\dt>0$, when we have
\be
\lambda_\omega={1\over\dt}\left(1-e^{-i\omega\dt}\right)-E
=\lambda^C_\omega+\ord{\dt}.
\ee

What went wrong in the naive continuum limit? The high
frequency contributions to the functional determinant which are
essential in the case of a first order differential operator
are missed in $\lambda^C$. The strong sensitivity on the
details of regulating the action at the scale $\dt$ can be seen
by moving the variable $\psid_j$ in the first term of the exponent
in \eq{impi} by $\dt$. It requires a change of sign of the
kinetic energy and replacing the term handling
the excitation by its complex conjugate,
\be\label{impm}
I_I=\prod_{k=0}^N\int d\psid_kd\psi_k\Psi_f^*(\psi_{N+1})\Psi_i(\psid_0)
e^{\sum_{j=0}^N[\psid_{j-1}(\psi_j-\psi_{j-1})-E\dt\psid_{j-1}\psi_j]}.
\ee
Another modification appearing in this formula is that the wave
functions are $\Psi(\psi)=a+b\psi$ and the excited state is represented
by $\Psi(\psi)=\psi$.

The problem arises because the trajectories are supposed to have
continuous first derivatives in classical calculus, behind the manipulations
leading from \eq{impi} to \eq{nimt}. This assumption is incorrect
in quantum physics. It is well known that the Brownian motion
trajectories are nowhere differentiable. Such
a singular structure is natural in the path integral solution
of the diffusion problem,
\be\label{brown}
P(X,T)={\cal N}^{-1}\lim_{N\to\infty}\prod_{j=1}^N\int dx_j
e^{-{N\over4DT}\sum_k(x_j-x_{j-1})^2}
\ee
since $N^2(x_j-x_{j-1})^2/T^2\approx2DN/T\to\infty$ for the typical
trajectories \cite{brown}.
The real time analogy of this phenomenon is that the trajectories
in the path integral representation for the solution of the
Schr\"odinger equation are nowhere differentiable.
This problem becomes more serious when the order of the differential
equation, the power of the velocity in the exponent of the path integral
is decreased, cf. \eq{impi}, \eq{brown}. Another source of concern
is the formal nature of the Grassmann variables. They possess no
numerical values and obey no concept of smoothness. These are the actual
sources of the difference between \eq{impi} and \eq{impm}.

The operator $G$ introduced in Eq. \eq{cggf} will be
replaced by the regulated expression
\be\label{gdefr}
G^{-1}_{x,n,x',n'}=\left[{1\over\dt}\left(\delta_{n,n'}-\delta_{n,n'+1}\right)
+\delta_{n,n'+1}\left(-{\hbar^2\over2m}\Delta_x-\mu+\ve_x\right)\right]
\delta_{x,x'},
\ee
where the notation $t=n\dt$ is used. In this manner the first term in the
right hand side of Eq. \eq{hfee} is obtained. The loop integrals containing the
propagators obtained from \eq{gdefr} have no more unexpected cut-off
effects and reproduce the usual continuum expressions for time scales
longer than $\dt$. As a result we shall deliberately use the continuum
propagators given by Eq. \eq{fgrsp} in our computation.

\section{Legendre transformation}\label{legend}
When the Legendre transform  is sought it might be advantageous
to convert the problem into those of solving a differential equation.
This analogy which gives the solution of Eq. \eq{dfltr}
in the simplest manner is discussed in this Appendix.

\subsection{Legendre transformation and differential equations}
Consider the non-linear differential equation $f(y')-ty'+y=0$
for the function $y(t)$ and the initial condition $y'(t_0)=v_0$
where $f'(v_0)=t_0$ and $f(x)$ is a convex function, $f''(x)>0$.
It is obvious that the pair of functions $y(t)$ and $f(x)$ are
Legendre transforms of each others. As an example consider the
choice $f(x)=ax^2/2+bx+c$, $v_0=0$ which leads to
$ay'^2/2+(b-t)y'+y=c$ and $y'(b)=0$ and the solution
$y(t)=t^2/2a-tb/a+b^2/2a-c$.

In a similar manner, the Legendre transformation \eq{dfltr} can be
written as a functional partial differential equation and
initial condition,
\be\label{lmdf}
W\left[{1\over i}{\delta\Gamma[\rho]\over\delta\rho}\right]
-\rho\cdot{\delta\Gamma[\rho]\over\delta\rho}+\Gamma[\rho]=0,\ \
{\delta\Gamma[\rho_0]\over\delta\rho}=i\sigma_0,
\ee
where $\delta W[\sigma_0]/\delta\sigma=i\rho_0$.
The quadratic generator functional
$W[\sigma]=\hf\sigma\cdot A\cdot\sigma+B\cdot\sigma+C$
gives the functional partial differential equation
\be
\Gamma[\rho]=\hf{\delta\Gamma[\rho]\over\delta\rho}\cdot A\cdot
{\delta\Gamma[\rho]\over\delta\rho}
+(\rho+iB)\cdot{\delta\Gamma[\rho]\over\delta\rho}-C,\ \
{\delta\Gamma[\rho_0]\over\delta\rho}=0
\ee
having the solution
\be\label{negyzsz}
\Gamma[\rho]=-\hf\rho\cdot A^{-1}\cdot\rho-i\rho\cdot A^{-1}\cdot B
+\hf B\cdot A^{-1}\cdot B-C.
\ee

\subsection{Non-interacting electrons}\label{lnonint}
The functional Cauchy problem of the effective action of the
non-interacting system is
\be
\Gamma^{\mathrm{free}}[\rho]
=-n_s\tr\log\left(G^{-1}-{\delta\Gamma^{\mathrm{free}}[\rho]
\over\delta\rho}\right)+\rho\cdot
{\delta\Gamma^{\mathrm{free}}[\rho]\over\delta\rho},\ \
{\delta\Gamma^{\mathrm{free}}[\rho]\over\delta\rho}_{\vert\rt}=0.
\ee
Expanding the logarithm and using the Taylor expansion \eq{vezsr} of $\rho$ in
powers of $\sigma$, one can rewrite the functional differential
equation
for $\Gamma^{\mathrm{free}}[\rho] $ in the form:
\bea\label{gfdiff}
\Gamma^{\mathrm{free}}[\rho]
=-n_s\tr\log G^{-1} -n_s \sum_{n=2}^\infty \biggl(1-{1\over n}\biggr)
 \int_{x_1,x_2,\ldots,x_n} S_{x_1,x_2,\ldots,x_n}
 {\delta \Gamma^{\mathrm{free}}[\rho]\over\delta\rho_{x_1} }
 {\delta \Gamma^{\mathrm{free}}[\rho]\over\delta\rho_{x_2} }\cdots
 {\delta \Gamma^{\mathrm{free}}[\rho]\over\delta\rho_{x_n} }.
\eea
Looking for the solution in the form of the functional Taylor
expansion \eq{gafree},
we can identify the expansion coefficients by taking the various functional
derivatives of both sides of Eq. \eq{gfdiff} in increasing
order. This yields just the result given in Eq. \eq{gaexfr}.
Thus, we recovered our previous result for the non-interacting
electron gas.

\subsection{Radiative corrections}
In the leading order one finds the functional (the right hand side of Eq.
\eq{gawei})
\be
W_0+\hf\left(\ei J\cdot\Delta-ie\rt\right)\cdot D\cdot
\left(\ei\Delta J-ie\rt\right)+{1\over 2e^2}J\cdot \Delta \cdot J
=W_0-{e^2\over2}\rt\cdot D\cdot\rt-i\rt\cdot D\cdot\Delta\cdot J
+\hf J\cdot\tG\cdot J
\ee
for $W[J]$ figuring in Eq. \eq{lmdf} with $J={1\over i}{\delta \Gamma[\rho]
\over\delta \rho}$
which, according to \eq{negyzsz}, reproduces \eq{densfm}.

\end{document}